\documentclass[final]{IEEEtran} % regular paper

\usepackage{amssymb,amsmath,amsthm}

\usepackage{graphicx}
\graphicspath{{figures-pdf/}}
\usepackage{url}

\usepackage{calc}   % Simple arithmetic in LATEX commands

\newlength\OneImW
\newlength\TwoImW
\newlength\FiveImW
\newlength\BigOneImW
\newlength\Onefigwidth
\newlength\figsep
\newtheorem{theorem}{Theorem}
\newtheorem{fact}{Fact}
\newtheorem{Property}{Property}
\newtheorem{Corollary}{Corollary}

\usepackage{hyperref}
\hypersetup{
	linktocpage=true,
	pdfborderstyle={/S/S/W 1},
	hyperindex=true,
	bookmarks=true,
	bookmarksopen=true,
	bookmarksnumbered=true,
}
\begin{document}
	
	\title{Dynamic Analysis of Digital Chaotic Maps via State-Mapping Networks}
	
	\author{Chengqing Li,~\IEEEmembership{Senior Member,~IEEE,} Bingbing Feng,
		Shujun Li,~\IEEEmembership{Senior Member,~IEEE,} J\"urgen Kurths,
		Guanrong Chen,~\IEEEmembership{Life Fellow,~IEEE}
		% <-this % stops a space
		\thanks{This work was supported by the National Natural Science Foundation of China (61772447, 61532020); DAAD/K.C.~Wong Fellowship (91664078); Royal Society, UK (IE111186); and the Hong Kong Research Grants Council under the GRF Grant CityU 11234916.}
		
		\thanks{C.~Li is with College of Computer Science and Electronic Engineering, Hunan University,
			Changsha 410082, Hunan, China (DrChengqingLi@gmail.com).}
		
		\thanks{B.~Feng is with College of Information Engineering, Xiangtan University, Xiangtan 411105, Hunan, China.}
		
		\thanks{S.~Li is with School of Computing \& Kent Interdisciplinary Research Centre in Cyber Security (KirCCS), University of Kent, Canterbury, Kent, CT2 7NF, UK.}
		
		\thanks{J.~Kurths is with Potsdam Institute for Climate Impact Research, Potsdam D-14415, Germany.}
		
		\thanks{G.~Chen is with Department of Electronic Engineering, City University of Hong Kong, Hong Kong SAR, China.}
	}
	
	% The paper headers
	\markboth{IEEE Transactions on Circuits and Systems I: Regular Papers}{Li\MakeLowercase{et al.}}
	
	% put a publisher's ID mark on the page
	\IEEEpubid{\begin{minipage}{\textwidth}\ \\[12pt] \centering
			1549-8328 \copyright 2019 IEEE. Personal use is permitted, but republication/redistribution requires IEEE permission.\\
			See http://www.ieee.org/publications standards/publications/rights/index.html for more information.
	\end{minipage}}
	
	\maketitle
	
	\begin{abstract}
		Chaotic dynamics is widely used to design pseudo-random number generators and for other applications such as secure communications and encryption. This paper aims to study the dynamics of discrete-time chaotic maps in the digital (i.e., finite-precision) domain. Differing from the traditional approaches treating a digital chaotic map as a black box with different explanations according to the test results of the output, the dynamical properties of such chaotic maps are first explored with a fixed-point arithmetic, using the Logistic map and the Tent map as two representative examples, from a new perspective with the corresponding state-mapping networks (SMNs). In an SMN, every possible value in the digital domain is considered as a node and the mapping relationship between any pair of nodes is a directed edge. The scale-free properties of the Logistic map's SMN are proved. The analytic results are further extended to the scenario of floating-point arithmetic and for other chaotic maps. Understanding the network structure of a chaotic map's SMN in digital computers can facilitate counteracting the undesirable degeneration of chaotic dynamics in finite-precision domains, helping also classify and improve the randomness of pseudo-random number sequences generated by iterating chaotic maps.
	\end{abstract}
	
	% https://www.sharelatex.com/learn/Nomenclatures
	% a system of naming things, especially in science
	\iffalse
	In IEEE journals, this is usually done as an unnumbered section before the introduction. You can use an IEEEdescription list with a little extra separation spacing via \IEEEusemathlabelsep:
	\section*{Nomenclature}
	\addcontentsline{toc}{section}{Nomenclature}
	\begin{IEEEdescription}[\IEEEusemathlabelsep\IEEEsetlabelwidth{$V_1,V_2,V_3$}]
		\item[$V_1,V_2,V_3$] Three-phase PWM output line voltages.
		\item[$\theta$] Rotor angle (in ``electrical degrees'').
		\item[$\omega$] Rotor (electrical) speed, corresponding to the time derivative of $\theta$.
	\end{IEEEdescription}
	\fi
	
	\section{Introduction}
	
	\IEEEPARstart{D}{ynamics} of chaos in the continuous (i.e., infinite-precision) domain is a fundamental topic in the fields of mathematical chaos theory and nonlinear sciences \cite{cqli:IEAIE:IE18}. Yet, the implementation of a chaotic system in a digital device is always an inevitable problem withholding its real applications \cite{kocarev2006discrete,cqli:autoblock:IEEEM18}. Under the joint influence of round-off errors and truncation errors (i.e., algorithmic errors) in a finite-precision domain (e.g. in a finite-state machine), a resultant digital orbit will be off-tracking from the theoretical one \cite{Oteo:error:PRE2007,Galias:rounding:CSM2013}. Based on the well-known \textit{shadowing lemma}, many believed that any pseudo-random number sequence generated by iterating a chaotic map retains the complex dynamics of the original chaotic map to a high extent \cite{Phatak:LogisticRNG:PRE95}. However, it was found that the dynamics of a digital chaotic map are definitely degraded to some degree \cite{Persohn:AnalyzeLogistic:CSF12}. In 1988, Yorke et al. investigated the period distribution of an orbit of the Ikeda map, starting from a specific initial point under various round-off precisions, and found that the expected number of periodic orbits is scaled to the precision \cite{Yorke:Round:PRA88}. In \cite{Li:DPWLCM:IJBC2005}, a set of objective metrics were proposed to measure the degree of dynamics degradation of piecewise-linear chaotic maps. In 1991, Chua \cite{lin1991chaos} suggested that a real digital filter can exhibit near-chaotic behaviors if its wordlength is sufficiently large (e.g., larger than 16 bits). 
	
	Due to the ``pigeonhole principle'' and a limited possible number of a digital state, the orbit generated by iterating
	a chaotic map from any initial state in the digital domain will definitely enter a periodic loop after the transient process,
	referring to the general \textit{cycle detection} problem about periodic functions, as discussed in
	\cite{sedgewick1982complexity,Philippe:Mapping:Crypto90}. The existence of a network relationship among all possible states of a given chaotic map in the digital domain was often ignored \cite{ozturk2014cycle}, instead having focus on statistics along
	the orbit (path) on the network. When the orbits of the Logistic map are computed in 64-bit floating-point precision,
	by statistical analysis it was shown in \cite{Oteo:error:PRE2007} how the errors change with respect to the control parameters. In \cite{Catchen2013period}, period distribution of the generalized discrete Arnold cat
	map was precisely derived. Furthermore, the maximum period of the sequences generated by iterating the Logistic map
	over the field $\mathbb{Z}_{3^n}$ was derived in \cite{liao:period:SCIS17}.
	In \cite{Takeru:Logistic:IEICE}, the influences of different rounding methods on the
	transient lengths and cycle lengths were analyzed experimentally. The influences caused by control parameters
	were further analyzed in \cite{Uehara:logistic:ITA}. In \cite{Persohn:AnalyzeLogistic:CSF12}, the periods and cycles
	of the Logistic map are exhaustively calculated in 32-bit floating point precision using high-throughput computing.
	
	\IEEEpubidadjcol % must call \IEEEpubidadjcol in the second column for its text to clear the %IEEEpubid mark.
	
	Some earlier works on the dynamics of digital chaos via studying the state-mapping network (or state transition diagram, or functional graph), composing of
	relationships between every pair of states, fall into the scope of network science studies. In 1986, Binder
	\cite{Binder:Logistic:PRA86} drew the state network of the Logistic map in the 5-bit fixed-point arithmetic domain
	and reported that the counterparts of some metrics for measuring dynamics in continuous chaos, such as
	Lyapunov exponent and entropy, work just as well. In \cite{Binder:cycles:PHyD1992}, he further experimentally
	studied how the number of limit cycles and the size of the longest cycle change with the fixed-point precision.
	Later, an analytical framework was proposed for recurrence network analysis of chaotic time series
	\cite{Kurths:PLA:2009,Donner:GeometryDynamics:EPJB2011}.
	In \cite{Shreim:NetworkCA:2007}, the complexity of 1-D cellular automata was classified by
	two parameters of its state-mapping network. However, the validity of such classification method was questioned in \cite{Cqli:CA:IJBC2017}. In \cite{Luque:FeigenbaumChaos:PLOS11}, an orbit of a state-mapping network (SMN) was transformed into a network via horizontal visibility. It was found that the network entropy can mimic the
	Lyapunov exponent of the original map in a subtle level.
	In \cite{Iba:LogisticNetwork:arxiv10,Borges:Mapping:EPJB07,Thurner:Logistic:ICCS2007,Iba:NetworkChaos:ICCS2011},
	a mapping network among sub-intervals was established to explore the coherence between network parameters
	and some well-recognized metrics characterizing chaotic dynamics. In \cite{xu:motif:PNAS08},  the relative
	frequency of different 4-node subgraphs of SMN of some chaotic maps and flows was
	used to discriminate the underlying chaotic systems.
	
	In retrospect, the seemingly complex dynamics of chaos has been very appealing for random number generation
	\cite{kocarev2003pseudorandom,Addabbo:Tent:ITIM2006,HPhu:DCS:SMCS2015}
	and random permutation \cite{masuda2002cryptosystems}. In fact, in 1947, von Neumann already suggested
	using the Logistic map as a pseudo-random number generator (PRNG) \cite{Neumann:logistic:BAMS47}.
	Since then, a large number of PRNG have been proposed based on various chaotic maps and their variants,
	e.g., the
	Logistic map \cite{heidari1994chaotic,Phatak:LogisticRNG:PRE95,Chen:random:CASII2010,LiCY:PRNS:VLSI2012,HPhu:DCS:SMCS2015},
	the Tent map \cite{jessa2002periodtent,masuda2002cryptosystems,Addabbo:Tent:ITIM2006}, the Sawtooth map
	\cite{jessa2006sawtooth,dastgheib2017digital}, the R{\'e}nyi chaotic map \cite{addabbo2007class:TCASI}, and the Cat map
	\cite{Catchen2013period}. In addition, chaos theory was widely used to design hash functions and encryption
	schemes. However, it is impossible for any chaotic map to reach the ideal chaotic state in a finite-precision digital domain. As reviewed in \cite{LiShujun:Rules:IJBC2006,Li:logistic:ND2014}, any dynamics degradation of digital chaos may facilitate thwarting security of the supporting encryption schemes.
	
	To counteract dynamics degradation, many methods have been proposed, for example adopting higher precision
	\cite{lin1991chaos}, perturbing chaotic states
	\cite{tao1998perturbance,heidari1994chaotic,Chen:random:CASII2010,LiCY:PRNS:VLSI2012},
	perturbing control parameters \cite{vcernak1996digital}, and cascading multiple chaotic
	maps \cite{Hua:model:CAS17}, switching multiple chaotic maps \cite{nagaraj2008increasing,YCZhou:Switching:TCASI2014},
	and feedback control \cite{addabbo2006feedback,HPhu:DCS:SMCS2015}. Most of these works claimed that the
	improved discrete chaotic maps can work as good alternatives of the classic PRNG, by showing that their results
	pass typical randomness test suites. The real structures of such chaotic maps implemented in the computer remain mysterious, in which
	some important details occurring with a very low probability were omitted by a limited number of random
	experiments. To improve such a situation, this paper studies the properties of SMN generated by iterating a
	chaotic map in the digital domain: every representable value in the domain of the chaotic map is
	considered as a node, while a directed edge between a pair of nodes is built if and only if the former node is
	mapped to the latter one by the chaotic map. Using the Logistic map and the Tent map as illustrative examples,
	the dynamical properties of chaotic maps in the fixed-point arithmetic domain are disclosed by studying their
	corresponding SMN. The scale-free properties of SMN are mathematically proved. The relationship between an SMN
	obtained in a floating-point arithmetic domain and that in a fixed-point arithmetic domain is revealed. Finally, it will
	be shown that SMN can work as fingerprints of chaos-based PRNG to coarsely evaluate their randomness.
	
	The remainder of the paper is organized as follows. Section~\ref{sec:networkFixed} performs network
	analysis on the SMNs of the Logistic map and the Tent map in the fixed-point arithmetic domain.
	Section~\ref{sec:networkFloat} presents an analysis of the SMNs of the two maps in the floating-point
	arithmetic domain. An application of SMN to the evaluation of PRNG is discussed in
	Sec.~\ref{sec:applyNetwork}. The last section concludes the investigation.
	
	\section{State-mapping network of digital chaotic maps in the fixed-point arithmetic domain}
	\label{sec:networkFixed}
	
	First, the state-mapping network of chaotic maps implemented in the fixed-point arithmetic precision is defined and its general properties are proved. Then, the special properties of SMN of the Logistic map and the Tent map are analyzed.
	
	\subsection{Basic properties of chaotic maps implemented in the fixed-point arithmetic precision}
	
	Given a map $f$: $[0, 1]\rightarrow [0, 1]$ and a computing domain of fixed-point arithmetic precision $n$, with a specified quantization scheme, its domain and range are both defined as a discrete set $\{x\ |\ x=\frac{i}{2^n}\}_{i=0}^{2^n}$. The associate \emph{state-mapping network} $F_n$
	is built in the following way: the $2^n+1$ possible states are viewed as $2^n+1$ nodes; every pair of
	nodes with labels $i$ and $j$ is linked with a directed edge if $f(i/2^n)=j/2^n$, calculated in the quantization domain. Let $f_n(i)$ denote
	the original value corresponding to the node with label $i$ in $F_n$ before the final quantization step,
	namely,
	\begin{equation*}
		F_n(i)=\mathrm{R}\left(f_n(i)\cdot 2^n\right),
	\end{equation*}
	where $\mathrm{R}(\cdot)$ is an integer quantization function, e.g. floor, ceil, and round. As discussed in \cite{Li:DPWLCM:IJBC2005}, a different quantization function only has a slight effect on the quantized value and does not significantly influence the overall structure of SMN. So, only round quantization is considered throughout the paper.
	
	Property~\ref{prop:evenrelation} below describes the relationship between the $i$-th node in $F_n$ and the $(2i)$-th node in $F_{n+1}$. Meanwhile, Properties~\ref{property:odd}, \ref{property:odd2} characterize that between the former and the $(2i\pm 1)$-th node in $F_{n+1}$.
	
	\begin{Property}
		\label{prop:evenrelation}
		The node with label ``$2i$'' in $F_{n+1}$ and that with label ``$i$'' in $F_{n}$ satisfy
		\begin{equation}
			F_{n+1}(2i)-2F_n(i)=
			\begin{cases}
				1          &  \text{if } r_n\in [0.25, 0.5); \\
				-1         &  \text{if } r_n\in [0.5, 0.75); \\
				0          &  \text{otherwise},
			\end{cases}
			\label{even_condition2}
		\end{equation}
		where
		\begin{equation*}
			r_n=\mathrm{frac}(f_{n}(i)\cdot 2^n),
		\end{equation*}
		$\mathrm{frac}(x)=x-\lfloor x\rfloor$, $i\in \{0, \cdots, 2^n\}$, and
		$\lfloor x\rfloor$ returns the largest integer less than or equal to $x$.
	\end{Property}
	\begin{IEEEproof}
		Since $F_{n+1}(2i)=\mathrm{R}\left(2\cdot f_{n+1}(2i)\cdot 2^{n}\right)$ and
		\begin{equation*}
			f_{n+1}(2i)\equiv f_n(i),
		\end{equation*}
		the proof of this property is straightforward using
		\begin{equation*}
			\mathrm{R}(2 x)=2\cdot \mathrm{R}(x)+
			\begin{cases}
				0   &  \mbox{if } 0    \le \mathrm{frac}(x)<0.25;   \\
				1   &  \mbox{if } 0.25 \le \mathrm{frac}(x)<0.5; \\
				-1  &  \mbox{if } 0.5  \le \mathrm{frac}(x)<0.75;\\
				0   &  \mbox{if } 0.75 \le \mathrm{frac}(x)\le 1.  \hspace{3em}\IEEEQEDhere
			\end{cases}
			\label{con:error} 
		\end{equation*}
	\end{IEEEproof}
	
	\begin{Property}
		\label{property:odd}
		The node with label ``$2i+1$'' in $F_{n+1}$ and that with label ``$i$'' in $F_{n}$ satisfy
		\begin{IEEEeqnarray}{rCl}
			\IEEEeqnarraymulticol{3}{l}{\left|F_{n+1}(2i+1)-2\cdot F_{n}(i)\right|}\nonumber\\
			\qquad & \le &    \left|\mathrm{R}\left( (f_{n+1}(2i+1)
			-f_{n+1}(2i))\cdot 2^{n+1} \right) \right|\nonumber\\
			&     &    \hspace{1cm}   \label{eq:oddCondition}
			\quad+\begin{cases}
				2  & \text{if } r_n\in [0.25, 0.75);\\
				1  & \text{otherwise},
			\end{cases}
		\end{IEEEeqnarray}
		and $i\in \{0, \cdots, 2^n-1\}$.
	\end{Property}
	\begin{IEEEproof}
		According to the triangular inequality, one has
		\begin{IEEEeqnarray}{rCl}
			\IEEEeqnarraymulticol{3}{l}{\left|F_{n+1}(2i+1)-2\cdot F_n(i)\right|\le} \nonumber\\
			\quad &      &   \left|F_{n+1}(2i+1)-F_{n+1}(2i)\right|+|F_{n+1}(2i)-2\cdot F_n(i)|.
			\label{eq:TriangleInequ}
		\end{IEEEeqnarray}
		Utilizing the property of the integer quantization function
		\begin{equation}
			\label{eq:integer}
			|\mathrm{R}(x)-\mathrm{R}(y)|\leq |\mathrm{R}(x-y)|+1,
		\end{equation}
		one obtains
		\begin{IEEEeqnarray}{rCl}
			\IEEEeqnarraymulticol{3}{l}{\left|F_{n+1}(2i+1)-F_{n+1}(2i)\right|} \nonumber \\
			\quad & =   &  \left| \mathrm{R}\left(f_{n+1}(2i+1)\cdot 2^{n+1}\right)
			-\mathrm{R}\left(f_{n+1}(2i)\cdot 2^{n+1}\right) \right| \nonumber \\
			\quad & \le &  \left| \mathrm{R}\left(f_{n+1}(2i+1)\cdot 2^{n+1}
			-f_{n+1}(2i)\cdot 2^{n+1}\right) \right| +1  \label{uppound}
		\end{IEEEeqnarray}
		Incorporating the above inequality and Eq.~(\ref{even_condition2}) into inequality~
		(\ref{eq:TriangleInequ}), the property is proved.
	\end{IEEEproof}
	
	\begin{Property}
		\label{property:odd2}
		The node with label ``$2i-1$'' in $F_{n+1}$ and that with label ``$i$'' in $F_{n}$ satisfy
		\begin{IEEEeqnarray}{rCl}
			\IEEEeqnarraymulticol{3}{l}{\left|F_{n+1}(2i-1)-2\cdot F_{n}(i)\right|}\nonumber\\
			\qquad & \le &    \left|\mathrm{R}\left( (f_{n+1}(2i-1)
			-f_{n+1}(2i))\cdot 2^{n+1} \right)\right|\nonumber\\
			&     &    \hspace{1cm}
			\quad+\begin{cases}
				2  & \text{if } r_n\in [0.25, 0.75);\\
				1  & \text{otherwise},
			\end{cases}
		\end{IEEEeqnarray}
		and $i\in \{1, \cdots, 2^n\}$.
	\end{Property}
	\begin{IEEEproof}
		As the proof is very similar to that of Property~\ref{property:odd}, it is omitted.
	\end{IEEEproof}
	
	\subsection{SMN of the digital Logistic map}
	
\begin{figure}[!htb]
		\centering
		\begin{minipage}{0.8\TwoImW}
			\centering
			\includegraphics[width=0.8\TwoImW]{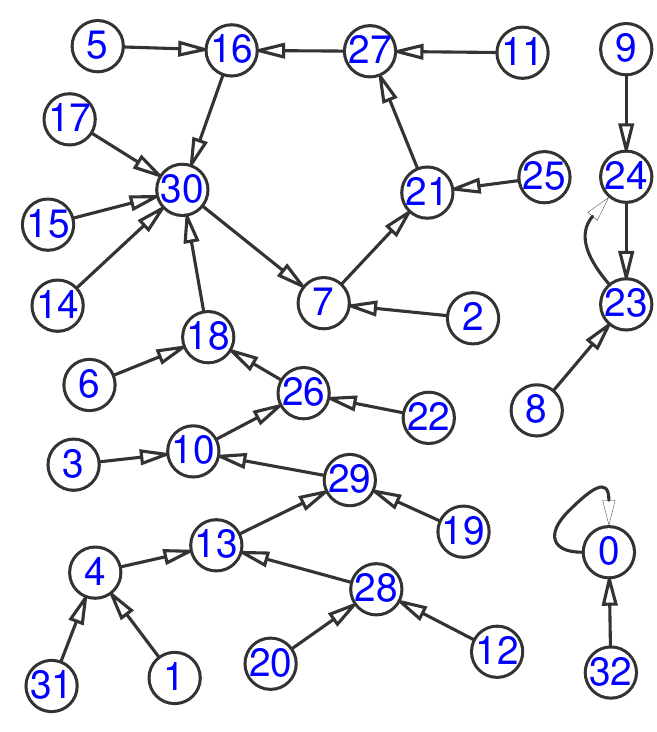}
			a)
		\end{minipage}\hspace{\figsep}
		\begin{minipage}{1.1\TwoImW}
			\centering
			\includegraphics[width=1.1\TwoImW]{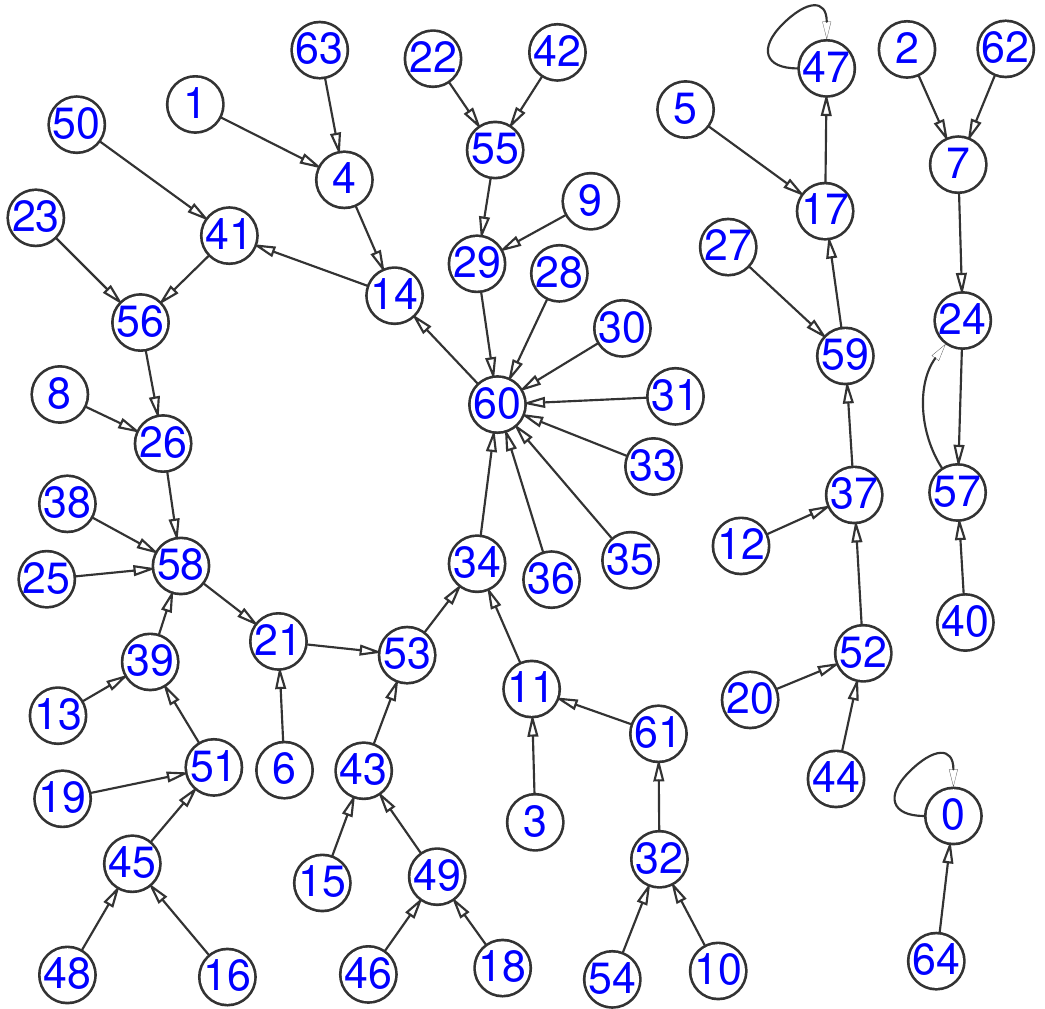}
			b)
		\end{minipage}\\
		\begin{minipage}{\Onefigwidth}
			\centering
			\includegraphics[width=\Onefigwidth]{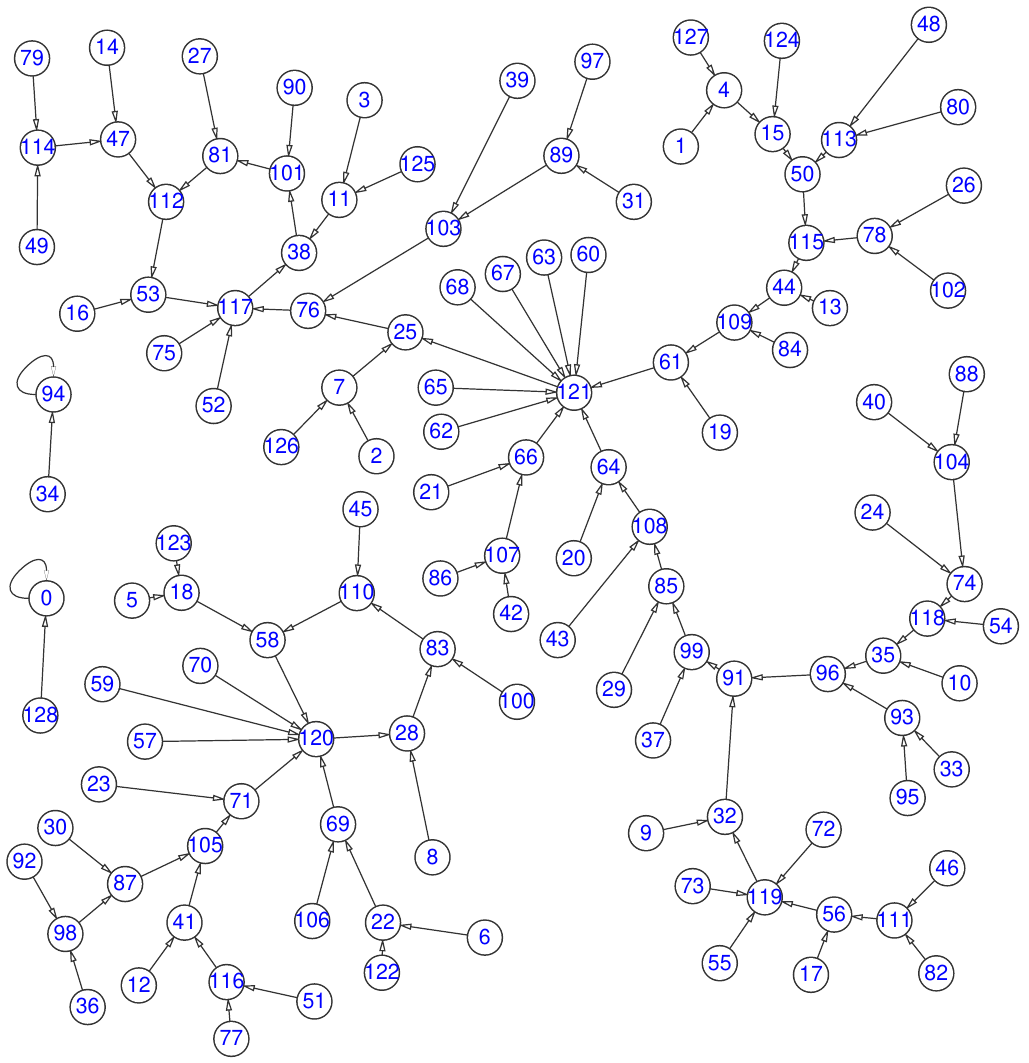}
			c)
		\end{minipage}
		\caption{SMNs of the digital Logistic map with $\mu=121/2^5$, implemented under different fixed-point precisions: a) $n=5$; b) $n=6$; c) $n=7$. (The first $2^n+1$ nodes are plotted.)}
		\label{fig:networkLogistic5and6bits}
	\end{figure}

	In the digital domain with fixed-point precision $n$, the Logistic map
	\begin{equation}
		f(x)=\mu\cdot x\cdot (1-x)
		\label{eq:logistic}
	\end{equation}
	becomes
	\begin{equation}
		f_n(i)=(N_{\mu}/2^{n_\mu})\cdot (i/2^n) \cdot (1-i/2^n),
		\label{eq:LogisticFinite}
	\end{equation}
	where $N_{\mu}$ is an odd integer in $\{0, \cdots, 2^{n_\mu+2}\}$, $\mu=N_{\mu}/2^{n_\mu}$,
	and $n_\mu \le n$. To facilitate the following discussion, draw the SMN of the Logistic map, $F^*_n$,
	with a fixed control parameter in the three arithmetic domains shown in
	Fig.~\ref{fig:networkLogistic5and6bits}a), b), and c), respectively.
	
	From Fig.~\ref{fig:networkLogistic5and6bits}, the following basic characteristics of SMN of the
	digital map can be noticed:
	\begin{itemize}[%
		\setlength{\labelwidth}{\widthof{\textbullet}}%
		\setlength{\labelsep}{3pt}%
		\setlength{\IEEElabelindent}{0pt}%
		\IEEEiedlabeljustifyl
		]
		
		\item The whole SMN is composed of a number of \textit{weakly connected components}, which are maximal subgraphs of a directed graph such that, for every pair of nodes $u$, $v$ in the  subgraph, there is a path between $u$ and $v$ in the underlying undirected version of the subgraph.
		
		\item Each weakly connected component has one and only one self-loop (an edge connecting a node
		to itself) or cycle (a sequence of nodes starting and ending at the same node such that, for every two
		consecutive nodes in the cycle, there exists an edge directed from the former node to the latter one.)
		
		\item Any node is linked to the cycle of the associated weakly connected component via a
		\textit{transient} process.
	\end{itemize}
	
	From Figs.~\ref{fig:networkLogistic5and6bits}a), b), c), one can further observe a special property of
	the SMN of the Logistic map: one weakly connected component dominates the whole SMN and there is
	a clear decreasing order among all the weakly connected components \cite{wang2004periodicity}.
	More precisely, the size of the component accounts for more than half of the size of the whole network.
	
	\begin{Property}
		\label{coro:logisticoddUpper}
		The nodes labeled number ``$i$'' in $F^*_{n}$ and that in $F^*_{n+1}$ with number ``$2i+1$''
		satisfy
		\begin{equation*}
			\left|F^*_{n+1}(2i+1)-2\cdot F^*_n(i)\right|\le
			\begin{cases}
				6  & \mbox{if } r_n \in [0.25, 0.75);\nonumber\\
				5  & \mbox{otherwise},
			\end{cases}
		\end{equation*}
		where $i\in \{0, \cdots, 2^n-1\}$, and $n\ge 3$.
	\end{Property}
	\begin{IEEEproof}
		By putting Eq.~\eqref{eq:LogisticFinite} into inequality~(\ref{eq:oddCondition}), one gets
		\begin{IEEEeqnarray*}{rCl}
			\IEEEeqnarraymulticol{3}{l}{\left|F^*_{n+1}(2i+1)-2\cdot F^*_n(i)\right|} \\
			\qquad & \le & \left|\mathrm{R}\left( \left(N_{\mu}/2^{n_\mu+2}\right)
			\cdot \left(4-(1+4i)/2^{n-1}\right)\right)\right|+ \\
			&     & \hspace{2cm}
			\begin{cases}
				2  & \mbox{if } r_n \in [0.25, 0.75); \\
				1  & \mbox{otherwise},
			\end{cases}
		\end{IEEEeqnarray*}
		based on Property~\ref{property:odd}.
		
		As $(N_{\mu}/2^{n_\mu+2})\in [0, 1]$, one furthermore has
		\begin{IEEEeqnarray}{rCl}
			\IEEEeqnarraymulticol{3}{l}{\left|F^*_{n+1}(2i+1)-2\cdot F^*_n(i)\right|}\nonumber\\
			\qquad & \le & \left|\mathrm{R}\left(4-(1+4i)/2^{n-1}\right)\right|+\nonumber\\
			\qquad &  & \hspace{4mm}
			\begin{cases}
				2  & \mbox{if } r_n \in [0.25, 0.75);\nonumber\\
				1  & \mbox{otherwise},
			\end{cases}\\
			\qquad & \le  &
			\begin{cases}
				6  & \mbox{if } r_n \in [0.25, 0.75);\nonumber\\
				5  & \mbox{otherwise}.   \hspace{14em}  \IEEEQEDhere
			\end{cases}
		\end{IEEEeqnarray}
	\end{IEEEproof}
	
\begin{figure}[!htb]
		\centering
		\begin{minipage}{\BigOneImW}
			\centering
			\includegraphics[width=\BigOneImW]{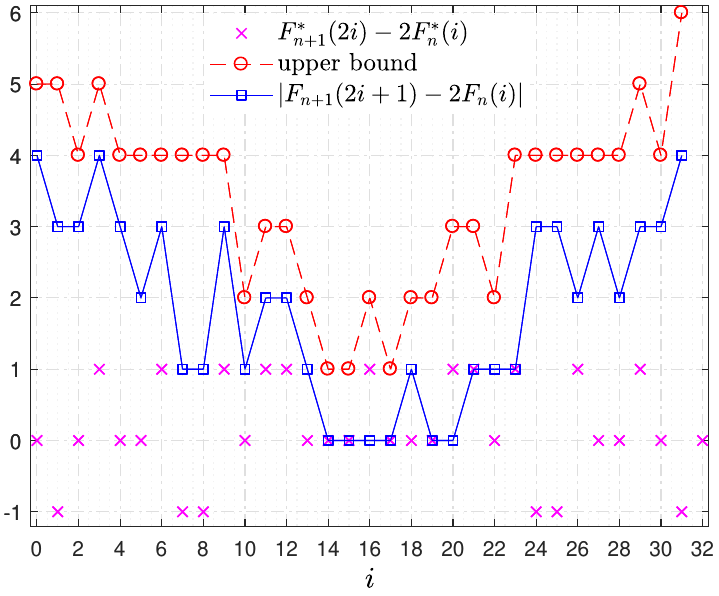}
			a)
			\hspace{1mm}
		\end{minipage}
		\begin{minipage}{\BigOneImW}
			\centering
			\includegraphics[width=\BigOneImW]{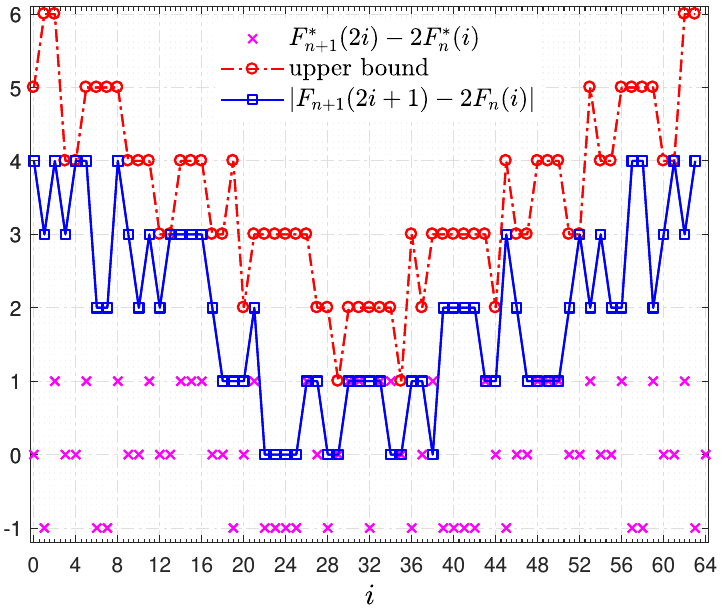}
			b)
		\end{minipage}
		\caption{Distribution of the differences between $F^*_n$ and $F^*_{n+1}$: a) $n=5$; b) $n=6$.}
		\label{fig:upperbound}
	\end{figure}

	From the proof of Property~\ref{coro:logisticoddUpper}, one can see that the upper bound of
	$|F^*_{n+1}(2i+1)-2\cdot F^*_n(i)|$ depends on the values of $i$ and $N_{\mu}$.
	Figure~\ref{fig:upperbound}a) depicts the values of $|F^*_{n+1}(2i+1)-2\cdot F^*_n(i)|$ and
	$F^*_{n+1}(2i)-2F^*_n(i)$ for every node shown in Fig.~\ref{fig:networkLogistic5and6bits}a).
	The corresponding data for the nodes shown in Fig.~\ref{fig:networkLogistic5and6bits}b) are
	plotted in Fig.~\ref{fig:upperbound}b), to further demonstrate the differences between $F^*_{n}$
	and $F^*_{n+1}$.
	
	\begin{fact}
		Round function $\mathrm{R}(\cdot)$ satisfies that
		$\mathrm{R}(x)=\mathrm{R}(y)$ if and only if
		\begin{equation*}
			\begin{cases}
				1/2\leq y-\lfloor x \rfloor<3/2  & \mbox{if } x-\lfloor x \rfloor\ge 1/2;\\
				-1/2\leq y-\lfloor x \rfloor<1/2  & \mbox{if } x-\lfloor x \rfloor<1/2,
			\end{cases}
		\end{equation*}
		where $x, y\in \mathbb{R}$.
		\label{fact:round}	
	\end{fact}
	
	\begin{Property}
		The in-degree of node $F^*_n(2^{n-1})$ in the SMN of the Logistic map implemented by $n$-bit finite precision is 
		\begin{equation}
			2\cdot \left\lfloor\sqrt{(d-\mathrm{R}(d)+1/2)\cdot 2^{n_\mu+n}/N_{\mu}}\right\rfloor+1,
		\end{equation}	
		where $d=N_{\mu}/2^{n_\mu-n+2}-\lfloor N_{\mu}/2^{n_\mu-n+2}\rfloor$.
		\label{prop:specialdegree}
	\end{Property}
	\begin{IEEEproof}
		Assuming that $F_n(i)=F_n(i+k_t)$, namely 
		\begin{equation}
			\mathrm{R}(f_n(i)\cdot 2^n)=\mathrm{R}(f_n(i+k_t)\cdot 2^n),
			\label{degreeCondition}
		\end{equation}
		one can ensure the degree of the node $F_n(i)$ among the SMN
		be $N_{k_t}$, where $k_t\in \mathbb{Z}$ and
		$N_{k_t}$ is the number of all possible values of $k_t$ satisfying Eq.~(\ref{degreeCondition}). 
		
		Putting Eq.~(\ref{eq:LogisticFinite}) with $i=2^{n-1}$ into Eq.~(\ref{degreeCondition}), one gets
		\begin{equation}
			\mathrm{R}(N_{\mu}/2^{n_\mu-n+2})=\mathrm{R}(N_{\mu}/2^{n_\mu-n} \cdot(1/4-k_t^2/2^{2n})).
			\label{nuu}
		\end{equation}
		Referring to Fact~\ref{fact:round}, one has
		\begin{equation}
			\mathrm{R}(d)-1/2\leq d-N_{\mu}k_t^2/2^{n_\mu+n}<\mathrm{R}(d)+1/2
			\label{inequality}
		\end{equation} 
		from Eq.~(\ref{nuu}), where one of the two cases in Fact~\ref{fact:round} is selected, depending on the value of $\mathrm{R}(d)\in \{0, 1\}$. Since $d-\mathrm{R}(d)<1/2$ for any $d$, the right part of the above inequality is always satisfied. Solving the left part of inequality (\ref{inequality}), one has 
		$|k_t|\leq \sqrt{(d-\mathrm{R}(d)+1/2)\cdot 2^{n_\mu+n}/N_{\mu}}$.
		So, $N_{k_t}=2\cdot \lfloor\sqrt{(d-\mathrm{R}(d)+1/2)\cdot 2^{n_\mu+n}/N_{\mu}}\rfloor+1$, 
		which completes the proof of the property.
	\end{IEEEproof}
	
	Since $f'(x)=\mu\cdot (1-2x)>0$ for $x\in [0, 1/2]$ and the value of the Logistic map monotonously
	increases from zero to the maximum value $f(1/2)=\mu/4$, which becomes $\mathrm{R}(N_{\mu}/2^{n_\mu-n+2})$ in the $n$-bit finite arithmetic domain. The point $1/2$ in the infinite-precision domain corresponds to the node $i=2^{n-1}$ in the SMN. When $n\neq n_\mu+1$,
	$N_{\mu}/2^{n_\mu-n+2}-\lfloor N_{\mu}/2^{n_\mu-n+2}\rfloor<1/2$ always holds, so
	$\mathrm{R}(N_{\mu}/2^{n_\mu-n+2})=\lfloor N_{\mu}/2^{n_\mu-n+2}\rfloor$.
	Since $f''(x)\equiv-2\mu<0$ for $x\in [0, 1/2]$, $f'(x)$ monotonously
	decreases from $\mu$ to zero, the node $F^*_n(2^{n-1})$ owns the maximal degree
	$
	2\cdot \left\lfloor\sqrt{ 2^{n_\mu+n-1}/N_{\mu}}\right\rfloor+1
	$
	in the associate SMN when $n>n_\mu+1$.
	
\begin{figure}[!htb]
		\centering
		\begin{minipage}{\BigOneImW}
			\centering
			\includegraphics[width=\BigOneImW]{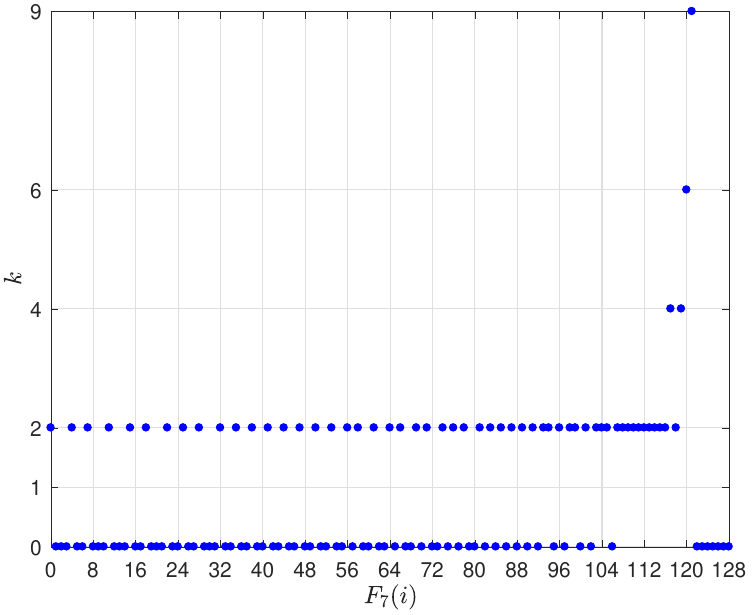}
			a)	
		\end{minipage}
		\begin{minipage}{\BigOneImW}
			\centering
			\includegraphics[width=\BigOneImW]{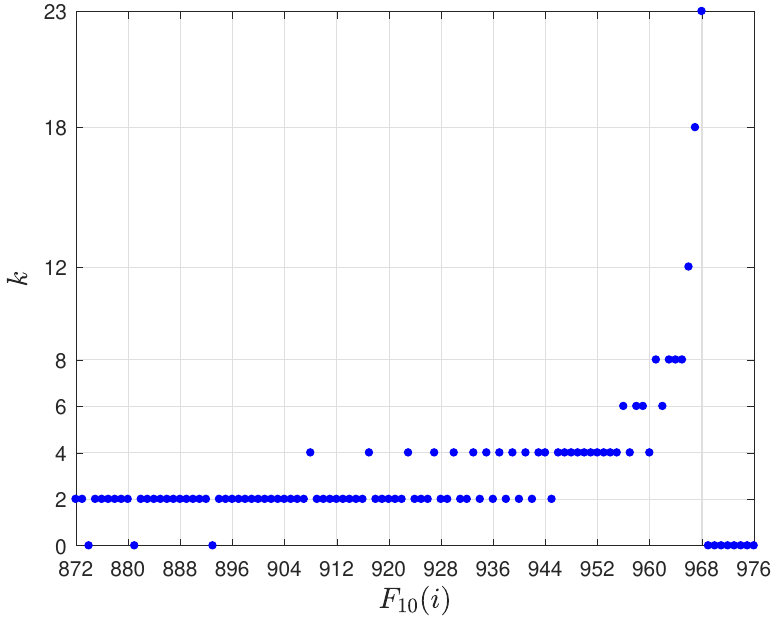}
			b)
		\end{minipage}
		\begin{minipage}{1.7\BigOneImW}
			\centering
			\includegraphics[width=1.7\BigOneImW]{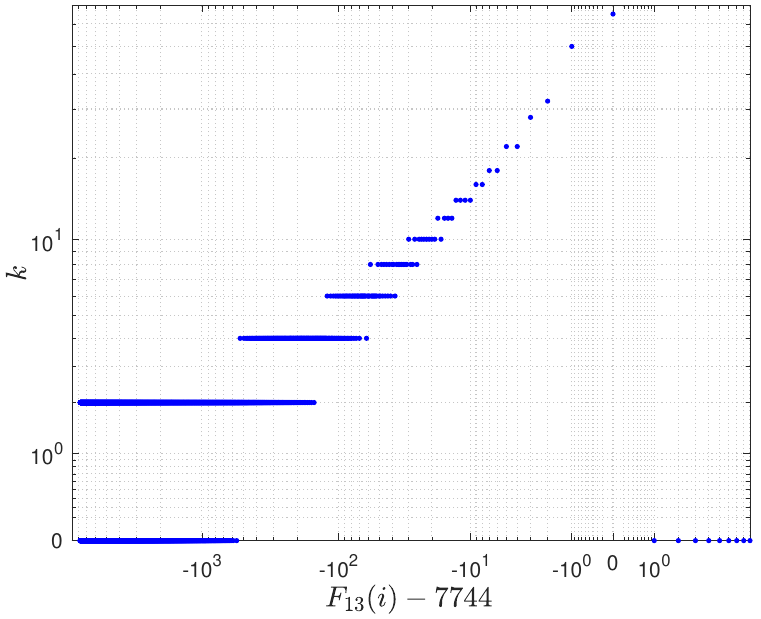}
			c)
		\end{minipage}
		\caption{The degree of $F^*_n(i)$ in the SMN of the Logistic map with $\mu=121/2^5$: a) $n=6$; b) $n=10$; c) $n=13$.}
		\label{fig:degreeIncrease}
	\end{figure}

	Since $f'(x)$ monotonously decreases from $\mu$ to zero in the studied interval,
	the in-degree (the number of edges directed into a node in a directed network) corresponding to $y$ monotonously
	increases as $y$ increases from zero to the maximum value $f(1/2)=\mu/4$.
	But, due to the quantization in the $n$-bit arithmetic domain, not every possible value in the codomain can be accessed by the digital version of the Logistic map (see Fig.~\ref{fig:degreeIncrease}). Fortunately, the quantization can only change the monotonicity when the degree is relatively small, as demonstrated in Fig.~\ref{fig:degreeIncrease}. As to the node $F^*_n(2^{n-1})$, its in-degree (the number of edges linking to the node) in the associate SMN has been exactly derived in Property~\ref{prop:specialdegree}. To obtain the overall relationship among the most important nodes in the SMN in terms of degrees, assume that the monotonicity is retained in the right part of the interval $(0, F^*_n(2^{n-1})/2^n)$
	in the following analysis. In-degree distribution $p(k)$, the fraction of nodes in the network with in-degree $k$, is a fundamental characteristic of a directed network. To derive the distribution of the SMN of the Logistic map, first compute its variant in Theorem~\ref{fig:degreeIncrease}, where cumulative in-degree distribution $P(k)$ means the fraction of nodes in the network with in-degrees larger than $k$.
	
	\begin{theorem}
		The cumulative in-degree distribution of the SMN $F^*_n$ approaches
		\[P(k)=\frac{4}{\mu^{2}k^{2}}\]
		as $n$ increases.
		\label{theorem:logisticMap}
	\end{theorem}
	\begin{proof}
		In a computing domain of fixed-point arithmetic precision $n$ with a specified quantization scheme,
		the domain and codomain of any 1-D map are both divided into intervals of fixed length
		$\Delta=1/2^n$.
		
		As demonstrated by Fig.~\ref{fig:degree}, assume a point $x_0$ and $y_0=f(x_0)$ are both multiples of $\Delta$.
		As for the interval to which $y_0$ belongs, the number
		of intervals having pre-image of $y_0$ in the neighborhood of $x_0$ is
		\begin{equation*}
			k=\left\lceil \frac{ |x_0-f^{-1}(f(x_0)-\Delta)| }{\Delta} \right\rceil,
		\end{equation*}
		where $\lceil \cdot \rceil $ gives the smallest integral value not less than the argument. 
		Since moving location of coordinate origin does not influence the value of the calculated degree,
		the above equation is applicable to any other point in the map.
		
		\begin{figure}[!htb]
			\centering
			\includegraphics[width=0.9\Onefigwidth]{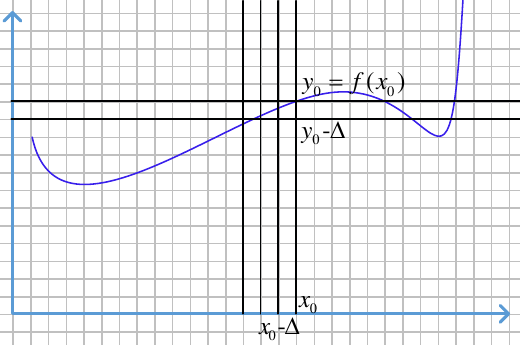}
			\caption{Demonstration of counting the number of preimages of a map in the digital domain.}
			\label{fig:degree}
		\end{figure}
		
		Since the Logistic map~(\ref{eq:logistic}) has a symmetric property, i.e.
		\begin{equation}
			f(x)=f(1-x),
			\label{eq:logisticsym}
		\end{equation}
		the in-degree of the node corresponding to $y=f(x)$ is double of that in the left part of the
		domain. In the interval $[0, 1/2]$, the inverse function of map~(\ref{eq:logistic}) is
		\begin{equation*}
			f^{-1}(y)=(1-\sqrt{1-4y/\mu})/2.
		\end{equation*}
		So, one has the in-degree of the node to which $y$ belongs, as
		\begin{IEEEeqnarray*}{rCl}
			k & = &  2 \cdot \left\lceil \frac{ f^{-1}(y)-f^{-1}(y-1/2^n)}{2^{-n}} \right\rceil \\
			& = &  2 \cdot \left\lceil \frac{\sqrt{1-4(y-1/2^n)/ \mu} -\sqrt{1-4y/ \mu} }{2^{1-n}} \right\rceil\\
			& = & 2 \cdot \frac{\sqrt{1-4(y-1/2^n)/ \mu} -\sqrt{1-4y/ \mu} }{2^{1-n}}+\epsilon,
		\end{IEEEeqnarray*}
		where $\epsilon$ is the change caused by the quantization function,
		and $0\le \epsilon<2$.
		Squaring both sides of the above equation twice, one gets
		\begin{equation}
			y=\frac{\mu}{4} -\frac{1}{\mu\cdot  (k-\epsilon)^2 }
			-\frac{\mu\cdot (k-\epsilon)^2}{2^{2n+4}} + 2^{-n-1}.
			\label{eq:expressy}
		\end{equation}
		As the relative influence of $\epsilon$ is very small, such similar cases are neglected in the following
		discussion.
		
		Since $f''(x)\equiv-2\mu<0$ for $x\in [0, 1/2]$, the rank (order) of the interval the state $y$ belongs, among all intervals, is
		\begin{equation*}
			r=\left\lceil \frac{\mu/4-y}{1/2^{n}} \right\rceil.
		\end{equation*}
		According to the definition of the cumulative in-degree distribution, one has
		\begin{IEEEeqnarray*}{rCl}
			P(k)&=&r/N\\
			&\approx &1-\frac{4y}{\mu},
		\end{IEEEeqnarray*}
		where 
		$N=\lceil \frac{\mu/4}{1/2^{n}}\rceil$ is the number of nodes in the SMN.
		Incorporating Eq.~\eqref{eq:expressy} into the above equation, one obtains
		\begin{equation*}
			P(k)=\left( \frac{2}{\mu k}- \frac{k}{2^{n+1}} \right)^2.
		\end{equation*}
		Obviously, $P(k)$ monotonously increases with respect to $n$. So, by increasing the value of $n$,
		the cumulative in-degree distribution of the SMN for the Logistic map tends to its limit:
		$\lim\limits_{n \to \infty}P(k)=\frac{4}{\mu^{2}k^{2}}$.
	\end{proof}
	
\begin{figure}[!htb]
		\centering
		\includegraphics[width=\Onefigwidth]{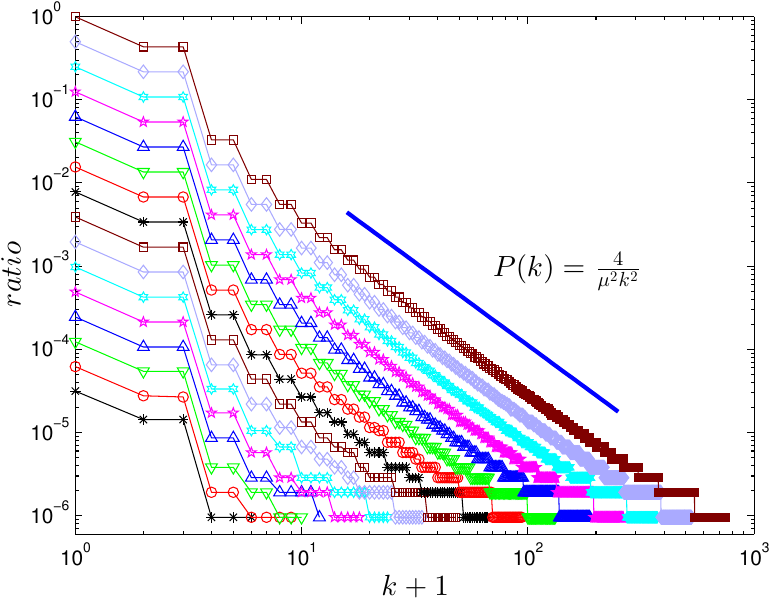}
		\caption{Cumulative in-degree distributions of SMN $F^*_{5} \sim F^*_{20}$, $\mu=\frac{121}{2^{5}}$.}
		\label{fig:CumulativeInDegreeDistribution_u121_n_5_20}
	\end{figure}

\begin{figure}[!htb]
		\centering
		\includegraphics[width=\Onefigwidth]{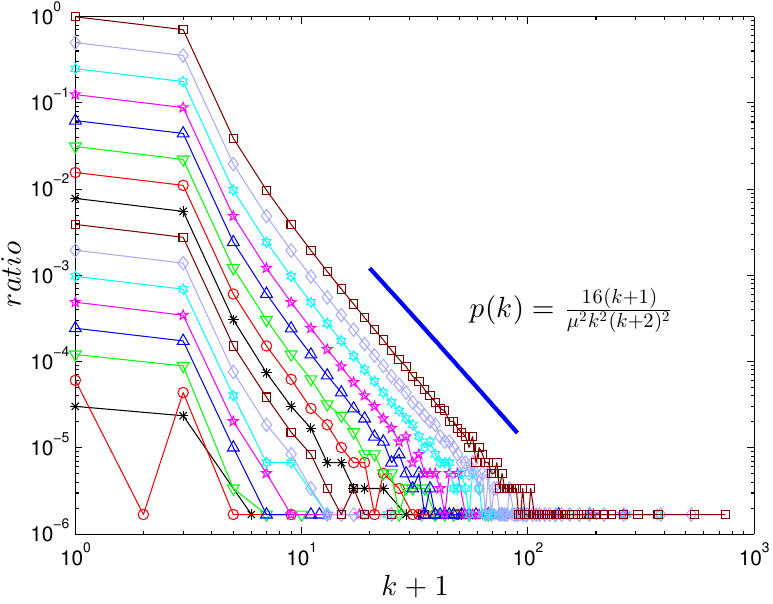}
		\caption{In-degree distributions of SMN $F^*_{5} \sim F^*_{20}$,
			$\mu=\frac{121}{2^{5}}$}
		\label{fig:InDegreeDistribution_u121_n_5_20}
	\end{figure}

	To verify Theorem~\ref{theorem:logisticMap}, draw the cumulative in-degree distributions of SMN $F^*_{5} \sim F^*_{20}$ as in
	Fig.~\ref{fig:CumulativeInDegreeDistribution_u121_n_5_20}, where $N$ is fixed to be $2^{20}$,
	to clearly demonstrate the evolution of the distributions. The corresponding in-degree distributions
	are shown in Fig.~\ref{fig:InDegreeDistribution_u121_n_5_20}, which agree with Corollary~\ref{coro:indegreedistribute}.
	
	\begin{Corollary}
		\label{coro:indegreedistribute}
		The in-degree distribution of the SMN $F^*_n$ satisfies
		\begin{equation*}
			p(k)=\frac{16(k+1)}{\mu^{2}k^{2}(k+2)^{2}}.
		\end{equation*}
	\end{Corollary}
	\begin{IEEEproof}
		Due to the symmetry property of the Logistic map, the in-degree of its SMN is always even,
		except the one corresponding to the critical point $f(1/2)$.
		
		According to the definition of the cumulative in-degree distribution, in-degree distribution
		$p(k)$ can be calculated by
		\begin{equation*}
			p(k)=P(k)-P(k+2).
		\end{equation*}
		So, one has
		\begin{IEEEeqnarray*}{rCl}
			p(k) & = & \left( \frac{2}{\mu k}- \frac{k}{2^{n+1}} \right)^2 - \left( \frac{2}{\mu (k+2)}
			- \frac{k+2}{2^{n+1}} \right)^2\\
			& = & (k+1) \left(\frac{16}{\mu^2 k^2 (k+2)^2} - \frac{1}{2^{2n}} \right).
		\end{IEEEeqnarray*}
		Obviously, $p(k)$ monotonously tends to its limit value:
		$\lim\limits_{n \to \infty} p(k)=\frac{16(k+1)}{\mu^2 k^2 (k+2)^2}$.
	\end{IEEEproof}
	
	\subsection{SMN of the digital Tent map}
	
\begin{figure}[!htb]
		\centering
		\begin{minipage}{\TwoImW}
			\centering
			\includegraphics[width=\TwoImW]{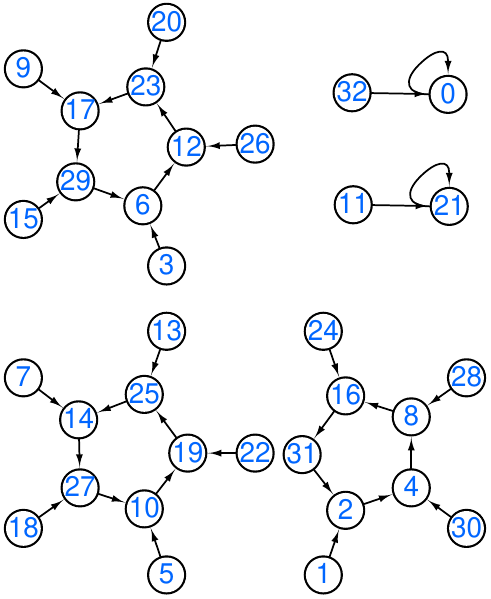}
			a)
		\end{minipage}
		\begin{minipage}{\TwoImW}
			\centering
			\includegraphics[width=\TwoImW]{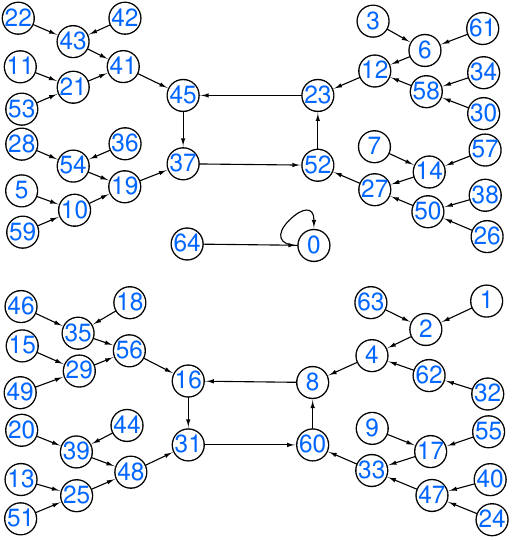}
			b)
		\end{minipage}
		\caption{SMN of the Tent map with $\mu=31/2^5$:
			a) 5-bit precision; b) 6-bit precision.}
		\label{fig:networkTent5and6bits}
	\end{figure}

	In the same digital domain discussed in the last subsection, the Tent map
	$f(x)=\mu \cdot\left(1-2\left|x-1/2\right|\right)$ becomes
	\begin{equation}
		f_n(i)=\left( (N_{\mu}/2^{n_\mu}) \cdot (1-2\left|(i/2^n)-1/2\right|) \right).
		\label{eq:TentFinite}
	\end{equation}
	To facilitate the following discussion, draw the SMN of the Tent map, $F^\star_n$, with
	$\mu=31/2^5$ in the domains of fixed-point 5-bit and 6-bit, respectively, as shown in
	Figs.~\ref{fig:networkTent5and6bits}a) and b).
	
	\begin{Corollary}
		\label{coro:tentoddUpper}
		The nodes with odd label numbers in the state network of $F^\star_{n+1}$ and that in the state
		network of $F^\star_{n}$ satisfy
		\begin{equation*}
			\left|F^\star_{n+1}(2i+1)-2\cdot F^\star_n(i)\right|\le
			\begin{cases}
				4  & \mbox{if } r_n \in [0.25, 0.75);\nonumber\\
				3  & \mbox{otherwise},
			\end{cases}
		\end{equation*}
		where $i\in \{0, \cdots, 2^n-1\}$.
	\end{Corollary}
	\begin{IEEEproof}
		The proof is very similar to that of Property~\ref{coro:logisticoddUpper}.
		Since $\left(N_{\mu}/2^{n_\mu-1}\right)\in[0,2]$, one has
		\begin{IEEEeqnarray}{rCl}
			\IEEEeqnarraymulticol{3}{l}{\left|F^\star_{n+1}(2i+1)-2\cdot F^\star_n(i)\right|}\nonumber\\
			\qquad & \le & \left|\mathrm{R}\left(N_{\mu}/2^{n_\mu-1}\right)\right|+\nonumber\\
			\qquad &     & \hspace{1cm}
			\begin{cases}
				2  & \mbox{if } r_n \in [0.25, 0.75);\nonumber\\
				1  & \mbox{otherwise},
			\end{cases}\\
			\qquad & \le  &
			\begin{cases}
				4  & \mbox{if } r_n \in [0.25, 0.75);\nonumber\\
				3  & \mbox{otherwise}. \hspace{14em}  \IEEEQEDhere
			\end{cases}
		\end{IEEEeqnarray}
	\end{IEEEproof}
	
\begin{figure}[!htb]
		\centering
		\begin{minipage}{\BigOneImW}
			\centering
			\includegraphics[width=\BigOneImW]{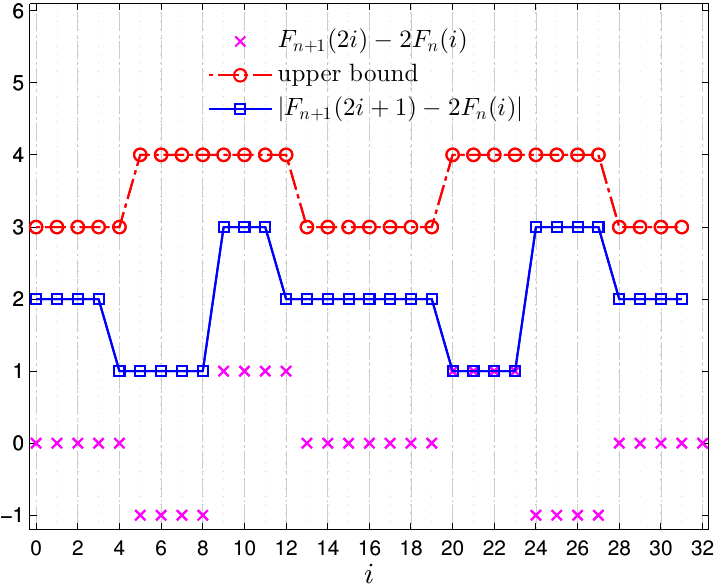}
			a)
		\end{minipage}
		\begin{minipage}{\BigOneImW}
			\centering
			\includegraphics[width=\BigOneImW]{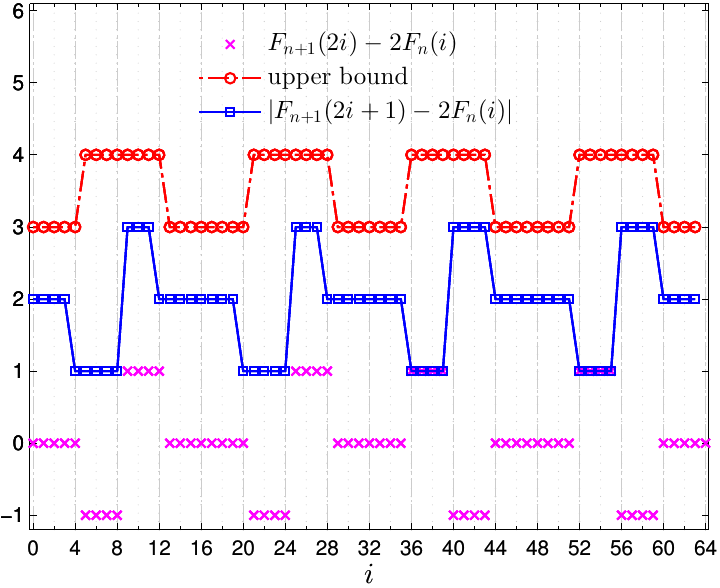}
			b)
		\end{minipage}
		\caption{Distributions of the differences between $F^\star_n$ and $F^\star_{n+1}$:
			a) $n=5$; b) $n=6$.}
		\label{fig:upperboundtent}
	\end{figure}

	From the proof of Corollary~\ref{coro:tentoddUpper}, one can see that the upper bound of
	$|F^\star_{n+1}(2i+1)-2\cdot F^\star_n(i)|$ depends only on $N_{\mu}$.
	Figure~\ref{fig:upperboundtent}a) depicts the values of $|F^\star_{n+1}(2i+1)-2\cdot
	F^\star_n(i)|$ and $F^\star_{n+1}(2i)-2F^\star_n(i)$ for every node shown in
	Fig.~\ref{fig:networkTent5and6bits}, which agrees with Properties~\ref{prop:evenrelation}
	and ~\ref{coro:logisticoddUpper}. The corresponding data with $n=6$ are shown in
	Fig.~\ref{fig:upperboundtent}b), which further demonstrate the differences between
	$F^\star_{n}$ and $F^\star_{n+1}$. Now, one can see that the SMN of the Tent map also incrementally expand, just as the Logistic map, in the same digital domain.
	
	\begin{figure}[!htb]
		\centering
		\begin{minipage}{\TwoImW}
			\centering
			\includegraphics[width=\TwoImW]{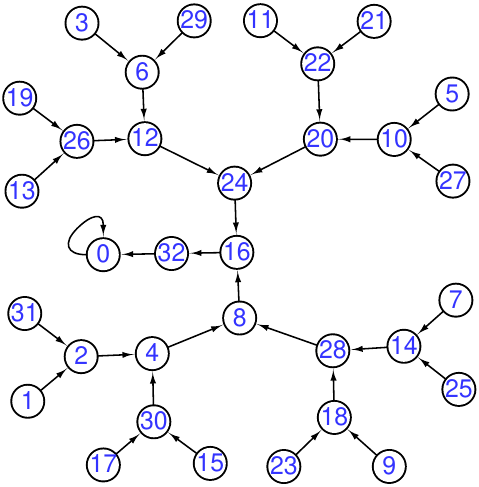}
			a)
		\end{minipage}
		\begin{minipage}{\TwoImW}
			\centering
			\includegraphics[width=\TwoImW]{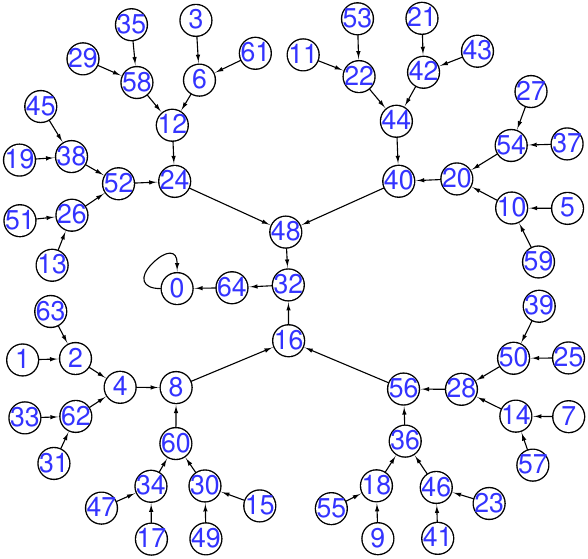}
			b)
		\end{minipage}
		\caption{SMN of the Tent map with $\mu=1$ after round quantization:
			a) 5-bit precision; b) 6-bit precision.}
		\label{fig:networktent5and6bitsu1}
	\end{figure}
	
	\begin{Property}
		\label{prop:degree}
		The degree of the SMN of the Tent map has only three possible values:
		\begin{equation*}
			k=
			\begin{cases}
				1  & \mbox{\rm the node denoting maximual value;}\\
				2  & \mbox{\rm other nodes owing pre-images;}\\
				0  & \mbox{\rm other nodes.}
			\end{cases}
		\end{equation*}
	\end{Property}
	\begin{proof}
		When $x=1/2$, the maximum value $f(x)=\mu$ is obtained. The node corresponding to
		$x=1/2$ only points to one node corresponding to $y=\mu$. So, the in-degree of the
		node to which $y=\mu$ belongs is $k=1$.
		
		Due to the symmetry property of the Tent map, only the left half part of the domain
		is considered. In the interval $[0, 1/2]$, the inverse function of the Tent map $y=f(x)$ is
		\begin{equation*}
			f^{-1}(y)=y/\left( 2\mu \right).
		\end{equation*}	
		Since $f'(x)=2\mu>0$ for $x\in [0,1/2]$, $f(x)$ monotonously increases with respect to $x$.
		Similarly to the proof of Theorem~\ref{theorem:logisticMap}, one has the in-degree of the node
		to which $y$ belongs, as
		\begin{IEEEeqnarray*}{rCl}
			k &=& 2\cdot \left\lceil \frac{f^{-1}(y)-f^{-1}(y-(1/2^{n}))}{2^{-n}} \right\rceil\\
			&=& 2\cdot \left\lceil \frac{y/(2\mu)-(y-1/2^{n})/(2\mu)}{2^{-n}} \right\rceil  \\
			&=& 2\cdot \left\lceil 1/(2\mu) \right\rceil\\
			&=& 2
		\end{IEEEeqnarray*}
		when $y \neq\mu$.
	\end{proof}
	
\begin{figure}[!htb]
		\centering
		\begin{minipage}{\OneImW}
			\centering
			\includegraphics[width=\OneImW]{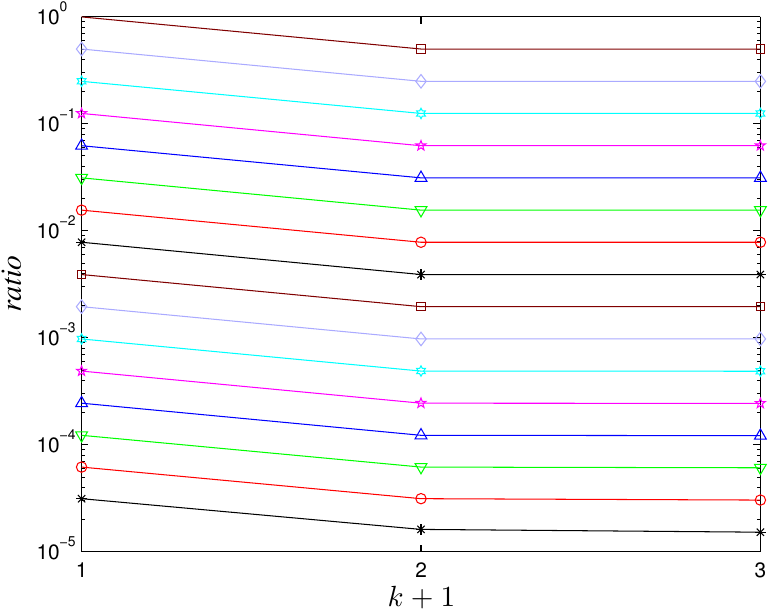}
			a)
		\end{minipage}
		\begin{minipage}{\OneImW}
			\centering
			\includegraphics[width=\OneImW]{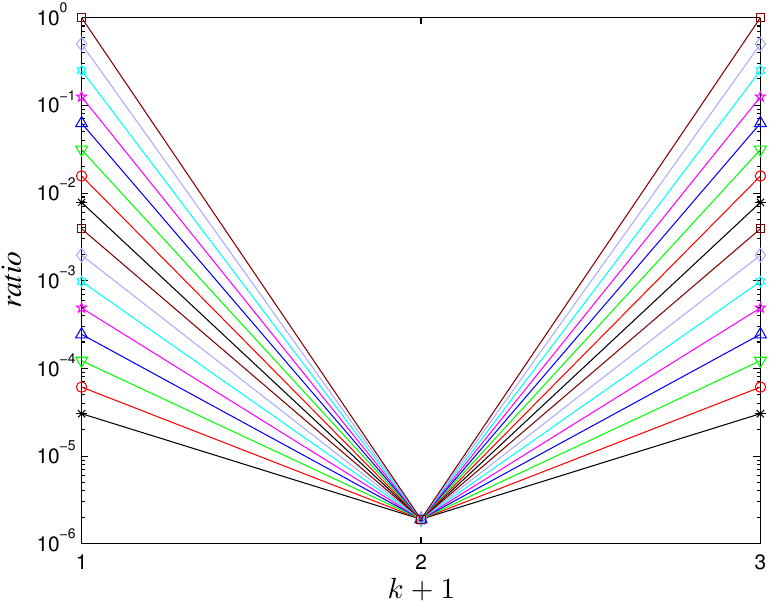}
			b)
		\end{minipage}
		\caption{Statistics of SMN $F^\star_{5} \sim F^\star_{20}$, $\mu=31/2^5$:
			a) accumulative in-degree distribution; b) in-degree distribution.}
		\label{fig:InDegreeDistribution3}
	\end{figure}

	From Property~\ref{prop:degree}, the edges in the SMN of the Tent map are
	not accumulated as the implementation precision increases (see Fig.~\ref{fig:InDegreeDistribution3}),
	which is different from that of the Logistic map.
	
	Based on the above discussions, one can conclude that $f''(x)>0$ in the whole domain is
	only a sufficient but not a necessary condition, for which the associate SMN of the corresponding map
	follows a power-law distribution.
	
	\section{Analysis of SMNs of digital chaotic maps in floating-point arithmetic domain}
	\label{sec:networkFloat}
	
	\subsection{Influence caused by floating-point arithmetic}
	
	To obtain a trade-off between a wide range and a high precision, digital computers adopt two kinds of binary representation formats to represent real numbers: fixed-point format and floating-point format. The
	former is more suitable for integers or real numbers with a fixed precision, and the latter for approximating
	real numbers with a higher and variable precision. Before standardization by IEEE and ANSI in 1985,
	different machines used dramatically different representation forms for the floating-point arithmetic.
	
	Following the floating-point standards, a sequence of $n$ bits, $\{b(i)\}_{i=0}^{n-1}$, is divided into
	three parts: a sign, a signed exponent, and a significand. The represented number is interpreted as the
	signed product of the significand and the number 2 to the power of its exponent:
	\begin{IEEEeqnarray*}{rCl}
		\IEEEeqnarraymulticol{3}{l}{v=}\\
		\begin{cases}
			0                                                                              & \text{if } e=0, \mathit{os}=0;        \\
			(-1)^{s}\cdot (\sum\limits_{i=1}^m b_{l+i}   \cdot 2^{-i})\cdot 2^{2-2^{l-1}}  & \text{if } e=0, \mathit{os}\neq 0;    \\
			(-1)^{s}\cdot \infty                                                           & \text{if } e=2^l-1, \mathit{os}=0;    \\
			\mbox{``not a number''}                                                  & \text{if } e=2^l-1, \mathit{os}\neq 0; \\
			(-1)^{s}\cdot \left(1+\sum\limits_{i=1}^{m} b_{l+i} \cdot 2^{-i} \right)\cdot 2^{e-\mathit{os}}
			& \text{otherwise},
		\end{cases}
	\end{IEEEeqnarray*}
	where $s=b_0$, $e=\sum\limits_{i=0}^{l-1} b_{1+i} \cdot 2^{i}$, $\mathit{os}=2^{l-1}-1$.
	%, ``NaN`` () is used to convey diagnostic information or escape-mechanism pointers.
	
	As to the single-precision floating-point format (binary32), e.g. ``float'' in C-language and
	``single'' in Matlab, $(l, m)=(8, 23)$, whereas $(l, m)=(11, 52)$ in double-precision floating-point
	format (binary64). In IEEE 754-2008, half-precision floating-point format (binary16) is designed for
	storaging with a higher precision, not for performing arithmetic computations, where $(l, m)=(5, 10)$.
	Meanwhile, due to the intermediate scale of the data generated by binary16, its simulation is widely
	adopted for experiments \cite{Rau:binary16}.
	
	In the floating-point arithmetic domain, equality (\ref{eq:logisticsym}) does not hold in general.
	Some concrete intermediate data from calculating the Logistic map are shown in
	Table~\ref{tab:differenceLogistic}. Specifically,
	\begin{equation*}
		%f(x)=\mu\cdot x\cdot (1-x)\stackrel{?}{=}f(1-x)=\mu \cdot(1-x)\cdot (1-(1-x)),
		\mu\cdot x\cdot (1-x)\stackrel{?}{=}\mu \cdot(1-x)\cdot (1-(1-x)),
	\end{equation*}
	which is caused by the difference between $\mathit{fl}(x)$ and $\mathit{fl}(1- \mathit{fl}(1-\mathit{fl}(x)))$,
	where $\mathit{fl}(x)$ denotes the normalized floating point number closest to $x$ in the given
	floating-point domain. If a number falls into the interval $[0.5, 1]$, its complement in terms of subtraction
	from 1 in that domain is fixed, so
	\begin{equation}
		\mathit{fl}(1- \mathit{fl}(1-\mathit{fl}(x)))\equiv
		\begin{cases}
			1- \mathit{fl}(1-\mathit{fl}(x))   &   \mbox{if } x\leq 0.5; \\
			\mathit{fl}(x)                     &   \mbox{if } x>0.5.
		\end{cases}
		\label{Condition:flxtwo}
	\end{equation}
	
	\begin{table*}[!htb]
		\centering
		\caption{Intermediate values of calculating the Logistic map in binary16.}
		\begin{tabular}{c *{4}{|c}}
			\hline
			$x$    &  $1-x$ & $1-(1-x)$  &   $f(x)$  &   $f(1-x)$ \\ \hline
			0.0099945068359375 & 0.98974609375 & 0.01025390625 & 0.037384033203125 & 0.038360595703125\\ \hline
			0.04998779296875 & 0.94970703125 & 0.05029296875 & 0.179443359375 & 0.1805419921875\\ \hline
			0.0899658203125  & 0.90966796875 & 0.09033203125 & 0.309326171875 & 0.310546875\\ \hline
			0.0999755859375  & 0.89990234375 & 0.10009765625 & 0.340087890625 & 0.340576171875\\ \hline
			0.199951171875   & 0.7998046875  & 0.2001953125  & 0.6044921875   & 0.60498046875\\ \hline
			0.289794921875   & 0.7099609375  & 0.2900390625  & 0.77783203125  & 0.7783203125\\ \hline
			0.389892578125   & 0.60986328125 & 0.39013671875 & 0.89892578125  & 0.8994140625\\ \hline
			0.489990234375   & 0.509765625   & 0.490234375   & 0.9443359375   & 0.94482421875\\ \hline
		\end{tabular}
		\label{tab:differenceLogistic}
	\end{table*}
	
	For any $x \in \mathbb{R}\cap [0, 1]$, there exists a unique integer $e$ such that
	$x=(\sum_{i=0}^\infty x_i\cdot 2^{-i})\cdot 2^e$, where $x_0=1$. In the floating-point domain
	with parameter $(l, m)$, it becomes
	\begin{equation*}
		\mathit{fl}(x)=
		\begin{cases}
			(\sum_{i=1}^{m} x_i\cdot 2^{-i})\cdot 2^{2-2^{l-1}}  &  \mbox{if } x\in (0,2^{2-2^{l-1}}); \\
			(1+\sum_{i=1}^{m} x_i\cdot 2^{-i})\cdot 2^e          &  \mbox{if } x\in [2^e,2^{e+1}),
		\end{cases}
	\end{equation*}
	where $e\in \{2-2^{l-1}, \cdots, -2\}$. Then, one has
	\begin{IEEEeqnarray}{lCl}
		1-\mathit{fl}(x)=\nonumber\\
		\quad\begin{cases}
			\sum_{i=1}^{2^{l-1}-2} 2^{-i}+(\sum_{i=1}^{m} \bar{x}_i\cdot 2^{-i}+ 2^{-m})\cdot 2^{2-2^{l-1}} & \\
			& \hspace{-3cm} \mbox{if } x\in (0,2^{2-2^{l-1}}); \\
			\sum_{i=1}^{-(e+1)}2^{-i}+(\sum_{i=1}^{m} \bar{x}_i\cdot 2^{-i}+ 2^{-m} )\cdot 2^{e}            & \\
			&  \hspace{-3cm} \mbox{if } x\in [2^e,2^{e+1}),
		\end{cases}
		\label{1minusflx}
	\end{IEEEeqnarray}
	where $\bar{x}_i=1-x_i$. Observing the first item in the right-hand side of Eq.~(\ref{1minusflx}), one can see
	that the exponent for the representation of $(1-\mathit{fl}(x))$ in that domain is minus one, namely,
	\[
	\mathit{fl}( 1-\mathit{fl}(x))= \left(1+\sum_{i=1}^m \hat{x}_i\cdot 2^{-i} \right)\cdot 2^{-1},
	\]
	where $\hat{x}_i\in\{0, 1\}$. In addition, $2^{l-1}-2\ge m+1$ is a necessary condition for ensuring
	a sufficient scope of represented numbers by the floating-point format. So, the two cases in
	Eq.~(\ref{1minusflx}) need to be further divided into three cases:
	\begin{IEEEeqnarray}{lCl}
		\mathit{fl}(1-\mathit{fl}(x))=\nonumber                                                                            \\
		\begin{cases}
			(\sum_{i=1}^{m+1} 2^{-i})               &   \hspace{-2.7cm}  \mbox{if } x\in (0,2^{2-2^{l-1}});   \\
			(\sum_{i=1}^{m+1} 2^{-i})                &   \hspace{-2.7cm}  \mbox{if } x\in [2^{e_1},2^{e_1+1}); \\
			(\sum_{i=1}^{-e_2-1} 2^{-i})+(\sum_{i=1}^{m+1+e_2} \bar{x}_i\cdot 2^{-i} )\cdot 2^{e_2}    &  \\
			&   \hspace{-2.7cm}  \mbox{if } x\in [2^{e_2},2^{e_2+1}),
		\end{cases}
		\label{fl1minusfx}
	\end{IEEEeqnarray}
	where $e_1\in \{2-2^{l-1}, \cdots, -m-2\}$, $e_2\in \{-m-1, \cdots, -2\}$.
	
	Referring to the first case in Eq.~(\ref{Condition:flxtwo}) and subtracting Eq.~(\ref{fl1minusfx})
	from Eq.~(\ref{1minusflx}), one obtains
	\begin{IEEEeqnarray}{lCl}
		\mathit{fl}(1-\mathit{fl}(1-\mathit{fl}(x)))-\mathit{fl}(x)     \nonumber\\
		= (1-\mathit{fl}(x))-\mathit{fl}(1-\mathit{fl}(x)) \nonumber\\
		=\begin{cases}
			(\sum_{i=m+2}^{2^{l-1}-2} 2^{-i})
			+(\sum_{i=1}^{m} \bar{x}_i\cdot 2^{-i}+ 2^{-m})\cdot 2^{2-2^{l-1}}   & \\
			&   \hspace{-3.1cm}  \mbox{if } x\in (0,2^{2-2^{l-1}}); \\
			(\sum_{i=m+2}^{-(e_1+1)} 2^{-i})+(\sum_{i=1}^{m} \bar{x}_i\cdot 2^{-i}+ 2^{-m} )\cdot 2^{e_1}  & \\
			&   \hspace{-3.1cm}  \mbox{if } x\in [2^{e_1},2^{e_1+1}); \\
			(\sum_{i=m+2+e_2}^{m} \bar{x}_i\cdot 2^{-i}+ 2^{-m} )\cdot 2^{e_2}      & \\
			&   \hspace{-3.1cm}  \mbox{if } x\in [2^{e_2},2^{e_2+1}).
		\end{cases}
		\label{eq:xdiffer1_x}
	\end{IEEEeqnarray}
	
\begin{figure}[!htb]
		\centering
		\includegraphics[width=\Onefigwidth]{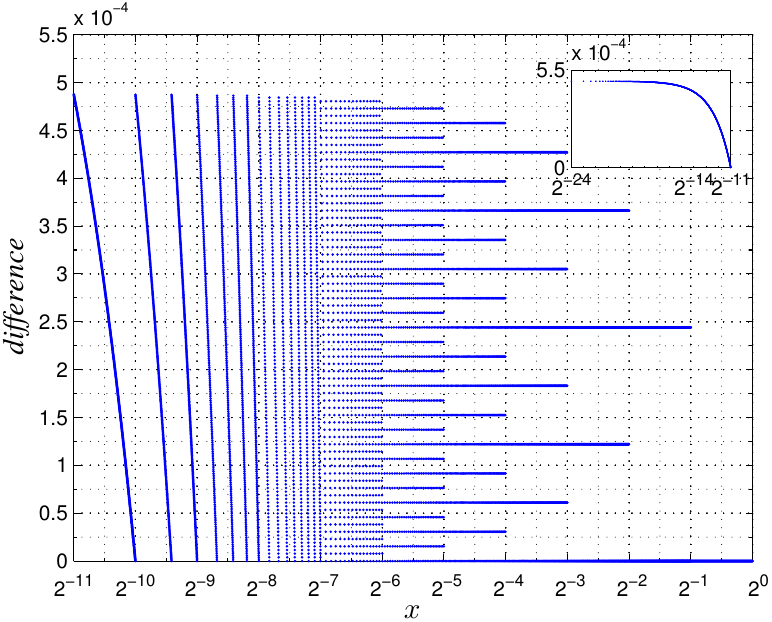}
		\caption{Subtracting $x$ from $1-(1-x)$ with various values of $x$ in Binary16.}
		\label{fig:differencex}
	\end{figure}

\begin{figure}[!htb]
		\centering
		\includegraphics[width=\Onefigwidth]{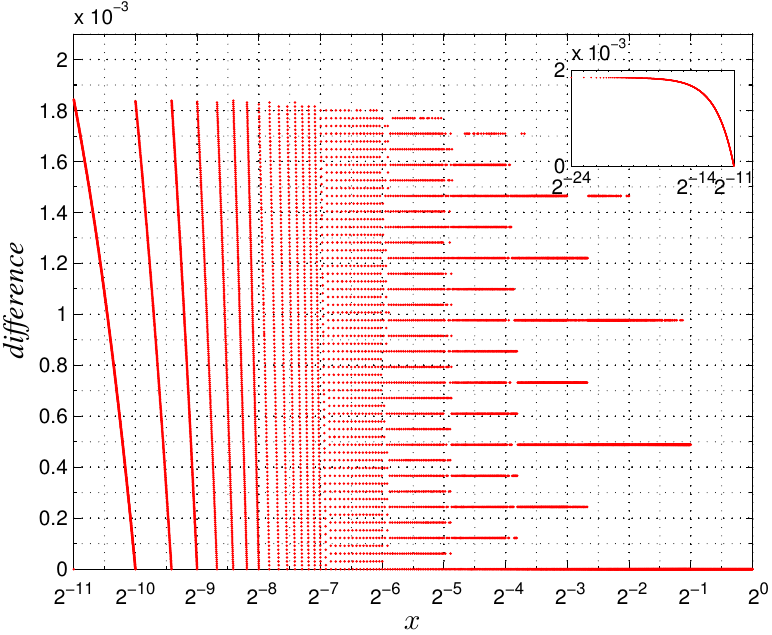}
		\caption{Subtracting $f(x)$ from $f(1-x)$ with various values of $x$ in Binary16.}
		\label{fig:differencefx}
	\end{figure}

	From Eq.~(\ref{eq:xdiffer1_x}), it follows that the difference between $1-(1-x)$ and $x$
	decreases monotonically in every selected interval, which is verified by the differences shown in
	Fig.~\ref{fig:differencex}. As shown in the inset in Fig.~\ref{fig:differencex}, the two segments
	corresponding the first two cases in Eq.~(\ref{eq:xdiffer1_x}), i.e. intervals
	$[2^{-10+2-2^4}, 2^{2-2^4}]=[2^{-24}, 2^{-14}]$ and
	$\{[2^{e_1}, 2^{e_1+1}]\}_{e_1=2-2^4}^{-10-2}
	=\{[2^{e_1}, 2^{e_1+1}]\}_{e_1=-14}^{-12}$, can be connected smoothly. The corresponding
	difference yielding the final result of the Logistic map is shown in Fig.~\ref{fig:differencefx}.
	
\begin{figure}[!htb]
		\centering
		\begin{minipage}{\TwoImW}
			\centering
			\includegraphics[width=\TwoImW]{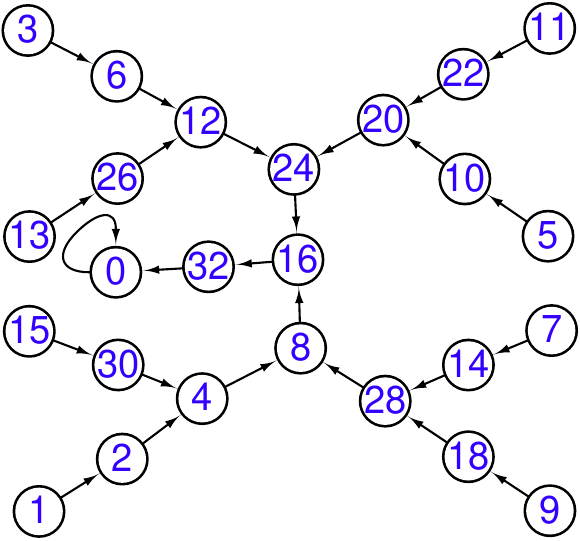}
			a)
		\end{minipage}
		\begin{minipage}{\TwoImW}
			\centering
			\includegraphics[width=\TwoImW]{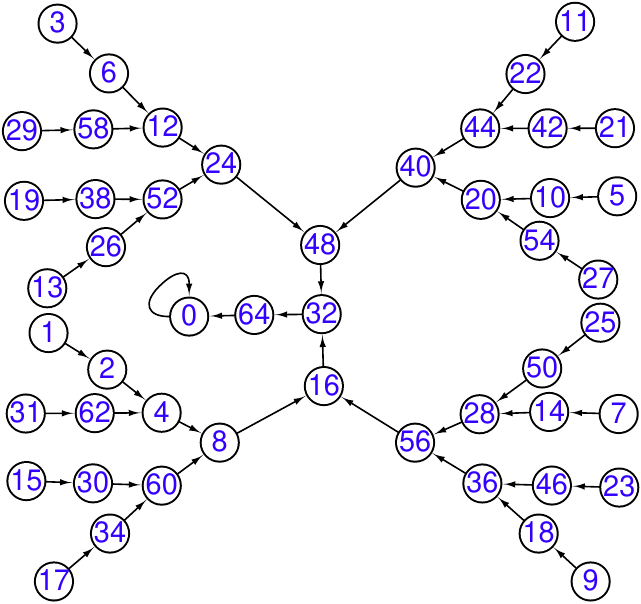}
			b)
		\end{minipage}
		\caption{SMN of the Tent map with $\mu=1$ in the binary floating-point domain:
			a) 7-bit binary floating-point domain ($l=3$, $m=3$);
			b) 8-bit binary floating-point domain ($l=3$, $m=4$).}
		\label{fig:networkTent7and8bitsu1}
	\end{figure}

	Assume that the initial condition is $x(0)=(0.b_1b_2\cdots b_j\cdots b_{L-1}b_L)_2\neq 0$,
	where $b_L=1$ (the least significant 1-bit) and $1-x(0)=(0.b_1'b_2'b_3'\cdots b_j'
	\cdots b_{L-1}'b_L)_2$. Then, the iteration of the Tent map becomes
	\begin{equation*}
		x(1)=\begin{cases} 2x(0)=x(0)\ll 1=(0.b_2\cdots b_j\cdots b_{L-1}b_L)_2, \\
			\hspace{4.4cm}                         \mbox{if }   0\leq x(0)<0.5,\\
			2(1-x(0))=(b_1'.b_2'b_3'\cdots b_j'\cdots b_{L-1}'b_L)_2,\\
			\hspace{4.4cm}                         \mbox{if } 0.5\leq x(0)\leq 1,
		\end{cases}
	\end{equation*}
	where $\ll$ denotes the left bit-shifting operation. Note that $b_1=0$ when $0\leq x(0)<0.5$.
	After $L-1$ iterations, one obtains $x(L-1)\equiv(0.b_L)_2=(0.1)_2$. So $x(L)\equiv 1$ and $x(L+1)\equiv 0$. That is, the number of required iterations converging
	to zero is $N_r=L+1$. Note that $N_r=0$ when $x(0)=0$. To visualize the operations of the
	digital Tent map with a typical example, the evolution process of the number is presented
	corresponding to the node labeled with ``$13$'' in Fig.~\ref{fig:networkTent7and8bitsu1}a):
	$(0.01101)_{2}\rightarrow (0.1101)_{2}\rightarrow (0.011)_{2}\rightarrow
	(0.11)_{2}\rightarrow (0.1)_{2}\rightarrow (1)_{2}\rightarrow (0)_{2}$.
	
	From the above analysis, it is clear that no any quantization error is introduced into
	the digital chaotic iterations, because the chaotic iterations can be exactly carried
	out with the digital operation $\ll$. The value of $L$ can be estimated according
	to two different conditions of $x(0)\neq 0$:
	% removing natural indent
	\begin{itemize}[%
		\setlength{\labelwidth}{\widthof{\textbullet}}%
		\setlength{\labelsep}{3pt}%
		\setlength{\IEEElabelindent}{0pt}%
		\IEEEiedlabeljustifyl
		]
		\item $x(0)$ is a normalized number:
		$x(0)=(1.b_{m-1}\cdots b_{0})\times 2^{-e}
		=(0.\overbrace{0\cdots 0}^{e-1}1b_{m-1}\cdots b_{0})_{2}$.
		Assuming that the least 1-bit of $x(0)$ is $b_{i}=1$, one can immediately get
		$x(0)=(0.\overbrace{0\cdots 0}^{e-1}1\overbrace{b_{m-1}\cdots
			b_{i}}^{m-i}\overbrace{0\cdots 0}^{i})_{2}$
		and deduce $L=(e-1)+1+(m-i)=e+(m-i)$. Considering $e\in [1,2^{l-1}-2]$ and
		$i\in [0,m-1]$, one has $L\in [2, 2^{l-1}-2+m]$.
		
		\item $x(0)$ is a non-zero denormalized number:
		$x(0)=(0.b_{m-1}\cdots b_{0})\times 2^{2-2^{l-1}}
		=(0.\overbrace{0\cdots 0}^{2^{l-1}-2}b_{m-1}\cdots b_{0})_{2}$.
		Assuming that the least 1-bit of $x(0)$ is $b_{i}=1$, one can immediately get
		$x(0)=(0.\overbrace{0\cdots 0}^{2^{l-1}-2}\overbrace{b_{m-1}\cdots
			b_{i}}^{m-i}\overbrace{0\cdots 0}^{i})_{2}$
		and deduce $L=2^{l-1}-2+(m-i)=2^{l-1}-2+m-i$. Considering $i\in [0,m-1]$, one has
		$L\in [2^{l-1}-1,2^{l-1}-2+m]$.
	\end{itemize}
	Summarizing, in both conditions, $L\le 2^{l-1}-2+m$.
	
	Next, consider the mathematical expectation of $i\in \{0,\cdots ,m-1\}$. Without loss of
	generality, for a denormalized number or a normalized number with a fixed exponent $e$,
	assume that the mantissa fraction $(b_{m-1}\cdots b_{0})_{2}$ distributes uniformly
	over the discrete set $\{0,\cdots ,2^{m}-1\}$. Then, the probability of
	$(b_{i}=1,b_{i-1}=\cdots =b_{0}=0)$ is $\frac{2^{m-1-i}}{2^{m}}=\frac{1}{2^{i+1}}$,
	and the probability of $(b_{m-1}=\cdots =b_{0}=0)$ is $\frac{1}{2^{m}}$. Thus,
	the mathematical expectation of $i$ is
	\begin{IEEEeqnarray*}{rCl}
		E(i) & \approx & \sum_{i=0}^{m-1} i\cdot \frac{1}{2^{i+1}} + m\cdot \frac{1}{2^{m}}\\
		& =       & \frac{1}{2}\cdot \sum_{i=1}^{m-1} \frac{i}{2^{i}} + \frac{m}{2^{m}}\\
		%     & =       & \frac{1}{2}\cdot \left(2-\frac{m+1}{2^{m-1}}\right) + \frac{m}{2^{m}}\\
		& =       & 1 - \frac{1}{2^{m}}.
	\end{IEEEeqnarray*}
	Next, the mathematical expectation of $e\in \{1,\cdots ,2^{l-1}-2\}$ is analyzed. From the
	uniform distribution of $x(0)$ in the interval [0,1], it follows that the probability of the exponent
	$e$ is about $\mathit{Prob}[2^{-e}\le x < 2^{-(e-1)}]=2^{-e}$. Thus, the mathematical
	expectation of $e$ is
	\begin{equation*}
		E(e)\approx \sum_{e=1}^{2^{l-1}-2} \frac{e}{2^{e}}=2-\frac{2^{l-1}}{2^{2^{l-1}-2}}\approx 2.
	\end{equation*}
	From the above deductions, one can deduce that
	\begin{IEEEeqnarray}{rCl}
		\IEEEeqnarraymulticol{3}{l}{E(L)}\nonumber\\
		& =       & \mathit{Prob}[\mathrm{normalized\ numbers}]\cdot \left(E(e)+(m-E(i))\right)+\nonumber\\
		&         & \mathit{Prob}[\mathrm{denormalized\ numbers}]\cdot \left(2^{l-1}-2+m-E(i)\right)\nonumber\\
		& =       & \frac{2^{l-1}-2}{2^{l-1}-1}\cdot \left(E(e)+(m-E(i))\right)+\nonumber\\
		&         & \frac{1}{2^{l-1}-1}\cdot \left(2^{l-1}-2+m-E(i)\right)\nonumber\\
		& \!\approx\! & \frac{2^{l-1}-2}{2^{l-1}-1}\cdot (2+(m-1))\!+\! \frac{1}{2^{l-1}-1}\cdot (2^{l-1}-2+m-1)\nonumber\\
		%& =       & \frac{2^{l-1}-2}{2^{l-1}-1}\cdot (m+1)+ \frac{1}{2^{l-1}-1}\cdot (2^{l-1}+m-3)\nonumber\\
		& =       & \frac{(2^{l-1}-2)(m+2)+m-1}{2^{l-1}-1}.
	\end{IEEEeqnarray}
	
\begin{figure}[!htb]
		\centering
		\includegraphics[width=\Onefigwidth]{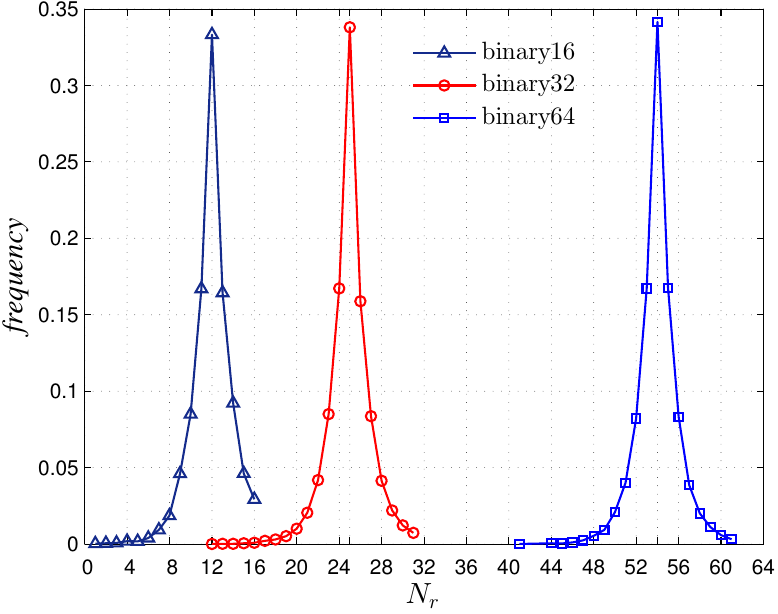}
		\caption{The occurrence frequency of different values of $N_r$ within a total of 10,000 values.}
		\label{fig:DistributionL_binary}
	\end{figure}

	To verify the mathematical expectation of $N_r$, some experiments were performed for testing
	on 10,000 initial conditions, pseudo-randomly generated with the standard Rand function of
	Microsoft Visual Studio 2010. The occurrence frequency of different values is shown in
	Fig.~\ref{fig:DistributionL_binary}, where the average values of $N_r$ for the three arithmetic
	domains are about 11.95, 24.97, and 54.01, respectively. All distributions well agree with the
	theoretical expectations.
	
	\subsection{Relationship between the SMNs obtained in two arithmetic domains}
	
	In the floating-point arithmetic domain with the exponent width of $l$ bits and the mantissa width of $m$ bits,
	the minimum fixed interval is $2^{( 1- (2^{l-1}-1) )} \cdot 2^{-m} = 2^{2-2^{l-1}-m}$. Under a
	fixed-point computing environment of precision $n$, the fixed interval is $2^{-n}$. If $2^{2-2^{l-1}-m}
	=2^{-n}$, i.e. $n=m+2^{l-1}-2$, the two SMNs obtained by implementing one map in the two
	domains have a strong correlation, as characterized by Theorem~\ref{theorem:fixedFloat}.
	
	\begin{theorem}
		Given binary floating-point format parameters $l$ and $m$, the node with label ``$i$'' in $F_{l,m}$ and
		that with label ``$i$'' in $F_{n}$ satisfy
		\begin{IEEEeqnarray}{rCl}
			\IEEEeqnarraymulticol{3}{l}{ F_{n}(i)- F_{l, m}(i) }\nonumber\\
			\qquad \le
			\begin{cases}
				1               &  \mbox{if } F_{n}(i)\in [0,2^{m});    \\
				2^{n-m-1-j}     &  \mbox{if } F_{n}(i)\in [2^{n-j-1},2^{n-j}),
			\end{cases}
		\end{IEEEeqnarray}
		where $j\in \{2^{l-1}-3, 2^{l-1}-4, \cdots, 1, 0\}$, and
		\begin{equation}
			\label{condition}
			n = m+2^{l-1}-2.
		\end{equation}
		\label{theorem:fixedFloat}
	\end{theorem}
	\begin{proof}
		Utilizing the property of the integer quantization function
		\begin{equation*}
			%\label{eq:xinteger}
			|x-\mathrm{R}(y)|= |\mathrm{R}(x-y)|,\ x\in \mathbb{Z},
		\end{equation*}
		one gets
		\begin{IEEEeqnarray*}{rCl}
			\IEEEeqnarraymulticol{3}{l}{\left|F_{l,m}(i)-F_n(i)\right|}\nonumber\\
			\qquad & = & \left|f_{l,m}(i) \cdot 2^{n}-\mathrm{R}\left(f_n(i) \cdot 2^{n}\right)\right|.\nonumber\\
			\qquad & = & \left| \mathrm{R}\left((f_{l,m}(i)-f_{n}(i))\cdot 2^{n}\right) \right|,
		\end{IEEEeqnarray*}
		where $F_{l, m}(i)=f_{l,m}(i) \cdot 2^{n}$.
		
		As for the node labeled with ``$i$'' in $F_{n}$, the corresponding value $i/2^n$ can be accurately
		represented in both the floating-point domain with $(l, m)$ and the $n$-bit fixed-point domain.
		Furthermore, the intermediate processes of calculating $f(i/2^n)$ are the same in the two arithmetic
		domains. So, the difference between $f_{l, m}(i)$ and $f_{n}(i)$ is caused only by the final
		quantization step.
		
		When $F_{n}(i)\in [2^{n-1-j}, 2^{n-j})$, one has
		\begin{align*}
			f_{l, m}(i)       = & 2^{-j-1}\cdot \left(1+\sum_{i=1}^m a_i \cdot 2^{-i} \right),\\
			f_{n}(i)          = & 2^{-j-1}\cdot \left(1+\sum_{i=1}^n c_i \cdot 2^{-i} \right),
		\end{align*}
		where $j\in \{0, \cdots, 2^{l-1}-3\}$. Obviously, $a_i=c_i$ for $i=1\sim m$. So, one has
		\begin{align*}
			f_{n}(i)-f_{l, m}(i) & = 2^{-j-1}\cdot \left( \sum_{i=m+1}^n c_i \cdot 2^{-i} \right) \\
			& < 2^{-m-j-1}.
		\end{align*}
		Thus, one can conclude that
		\begin{IEEEeqnarray*}{rCl}
			\IEEEeqnarraymulticol{3}{l}{F_n(i)- F_{l,m}(i) }\nonumber\\
			\qquad & =   & \mathrm{R}\left( ( f_{n}(i)-f_{l,m}(i) )\cdot 2^{m+1+j}
			\cdot 2^{n-m-1-j}\right)\nonumber\\
			\qquad & \le & \mathrm{R}\left(2^{n-m-1-j}\right)\nonumber\\
			\qquad & =   & 2^{n-m-1-j}.
		\end{IEEEeqnarray*}
		
		When $F_{n}(i)\in [0, 2^{m})$, $f_{l,m}(i)$ is a subnormal number, and $f_{l,m}(i)$ and $f_{n}(i)$
		can be expressed as
		\begin{align*}
			f_{l, m}(i)      & = 2^{2-2^{l-1}}\cdot \left( \sum_{i=1}^m a_i \cdot 2^{-i} \right),\\
			f_{n}(i)         & = 2^{2-2^{l-1}}\cdot \left( \sum_{i=1}^n c_i \cdot 2^{-i} \right).
		\end{align*}
		Similarly to the above cases, one has
		\begin{align*}
			f_{n}(i)-f_{l, m}(i) & =  2^{2-2^{l-1}}\cdot \left(\sum_{i=m+1}^n c_i \cdot 2^{-i}\right)\\
			& <  2^{-m+2-2^{l-1}} \\
			& =  2^{-n}.
		\end{align*}
		So, one gets
		\begin{align*}
			\left|F_{l,m}(i)-F_n(i)\right|  & =  \mathrm{R}\left(\left| (f_{l,m}(i)-f_{n}(i))\cdot 2^{n} \right|\right)\\
			& \le  \mathrm{R}\left(1\right)= 1.
		\end{align*}	
	\end{proof}
	
\begin{figure}[!htb]
		\centering
		\begin{minipage}{0.8\TwoImW}
			\centering
			\includegraphics[width=0.8\TwoImW]{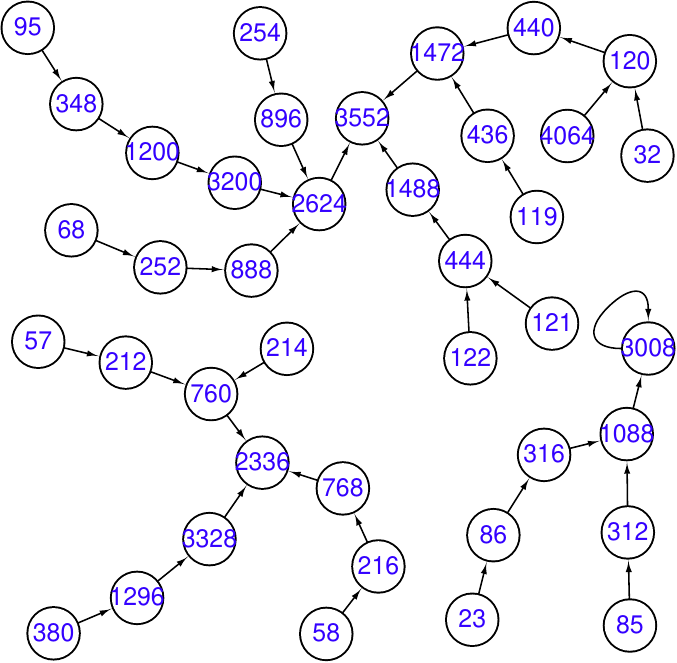}
			a)
		\end{minipage}
		\hspace{1mm}
		\begin{minipage}{1.15\TwoImW}
			\centering
			\includegraphics[width=1.15\TwoImW]{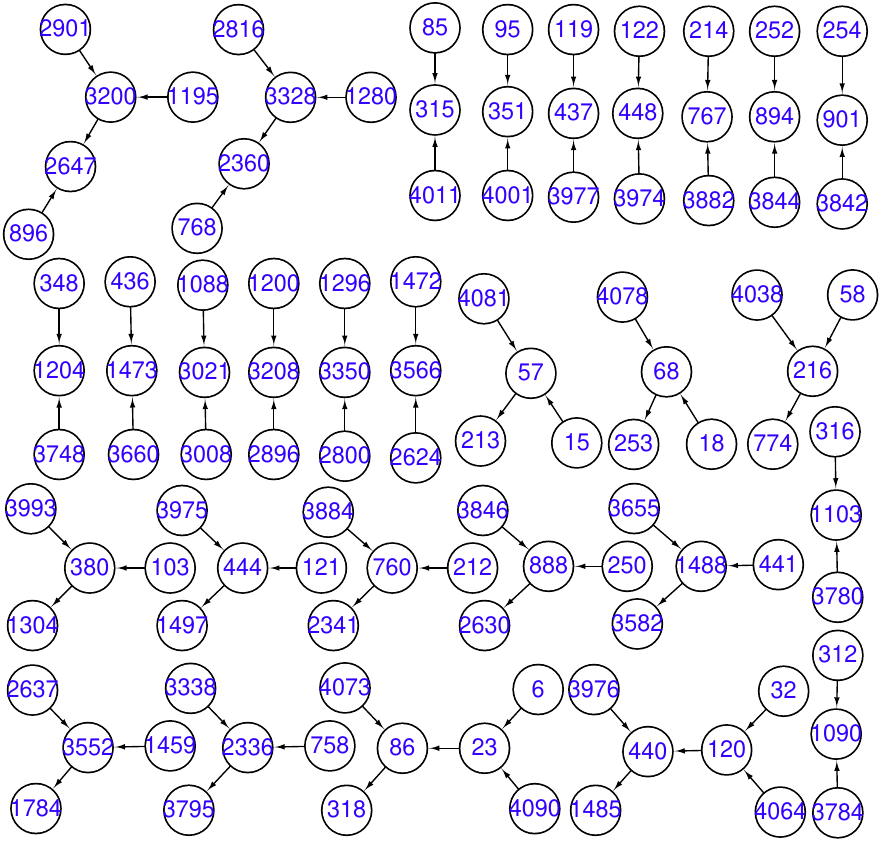}
			b)
		\end{minipage}
		\caption{Some connected components of the SMN of the Logistic map in floating-point domain
			and relative relations of their nodes in the corresponding fixed-point domain:
			a) 11-bit floating-point domain with ($l=4$, $m=6$); b) 12-bit finite-precision domain,
			where $\mu=121/2^5$.}
		\label{fig:networklogistic12and11bits}
	\end{figure}

	To illustrate Theorem~\ref{theorem:fixedFloat}, draw three connected components of the SMN of
	the Logistic map with $\mu=121/2^5$, $F_{12}$, in Fig.~\ref{fig:networklogistic12and11bits}a).
	Relative relations of the nodes in $F_{4, 6}$ are shown in Fig.~\ref{fig:networklogistic12and11bits}b).
	The corresponding parts of $F_{4, 6}$ are shown in Fig.~\ref{fig:networklogistic12and11bits}b).
	
	Due to the space limitation, only the involved nodes and their neighbors, but not the original
	connected components, are shown in Fig.~\ref{fig:networklogistic12and11bits}b). Moreover,
	differences between some mappings in $F_{12}$ and that in $F_{4, 6}$ are listed in
	Table~\ref{tab:difference}, which validates Theorem~\ref{theorem:fixedFloat} as well.
	
	\begin{table}[!htb]
		\centering
		\caption{Differences between SMNs implemented in two arithmetic domains with $n=12$, $l=4$, $m=6$.}
		\begin{tabular}{c *{4}{|c}}
			\hline
			$i$    &  $F_{n}(i)$ & $F_{l,m}(i)$  &   $\left|F_{n}(i)-F_{l,m}(i)\right|$ & $2^{n-m-1-j}$\\ \hline
			6    & 23   & 22   & 1  & 1  \\ \hline
			18   & 68   & 67   & 1  & 1  \\ \hline
			33   & 124  & 123  & 1  & 1  \\ \hline
			40   & 150  & 148  & 2  & 2  \\ \hline
			53   & 198  & 196  & 2  & 2  \\ \hline
			67   & 249  & 248  & 2  & 2  \\ \hline
			82   & 304  & 300  & 4  & 4  \\ \hline
			112  & 412  & 408  & 4  & 4  \\ \hline
			130  & 476  & 472  & 4  & 4  \\ \hline
			156  & 567  & 560  & 7  & 8  \\ \hline
			238  & 848  & 840  & 8  & 8  \\ \hline
			284  & 999  & 992  & 7  & 8  \\ \hline
			316  & 1103 & 1088 & 15 & 16 \\ \hline
			576  & 1872 & 1856 & 16 & 16 \\ \hline
			648  & 2063 & 2048 & 15 & 16 \\ \hline
			768  & 2360 & 2336 & 24 & 32 \\ \hline
			1280 & 3328 & 3296 & 32 & 32 \\ \hline
			2080 & 3871 & 3840 & 31 & 32 \\ \hline
		\end{tabular}
		\label{tab:difference}
	\end{table}
	
	From Theorem~\ref{theorem:fixedFloat}, one can see that the SMN of the Logistic map implemented by
	the floating-point arithmetic $(m, l)$ can be regarded as a rewired version of the sub-network of
	its SMN implemented in the corresponding fixed-point precision. More precisely, SMN $F_{l,m}$ can
	be generated from SMN $F_{n}$ as follows: all nodes linking to $F_{n}(i)$ are redirected to the nodes labeled $2^{n-j}-(k+1)\cdot 2^{n-m-1-j}$
	when $2^{n-j}-(k+1)\cdot 2^{n-m-1-j} < F_{n}(i) < 2^{n-j}-k\cdot 2^{n-m-1-j}$; all nodes linking
	to $F_{n}(i)$, except the node labeled $2^{n}$, are redirected to the nodes with label $2^{n-j}
	-k\cdot 2^{n-m-1-j}$ or $2^{n-j}-(k+1)\cdot 2^{n-m-1-j}$ when $F_{n}(i)= 2^{n-j}-k\cdot
	2^{n-m-1-j}$, where $j\in \{2^{l-1}-4, \cdots, 1, 0\}$, and $k=0\sim 2^{m}-1$. When $0 < F_{n}(i) < 2^{m+1}$, all nodes linking to $F_{n}(i)$ are redirected to the nodes with label $F_{n}(i)$ or $F_{n}(i)-1$.
	
\begin{figure}[!htb]
		\centering
		\begin{minipage}{0.9\TwoImW}
			\centering
			\includegraphics[width=0.9\TwoImW]{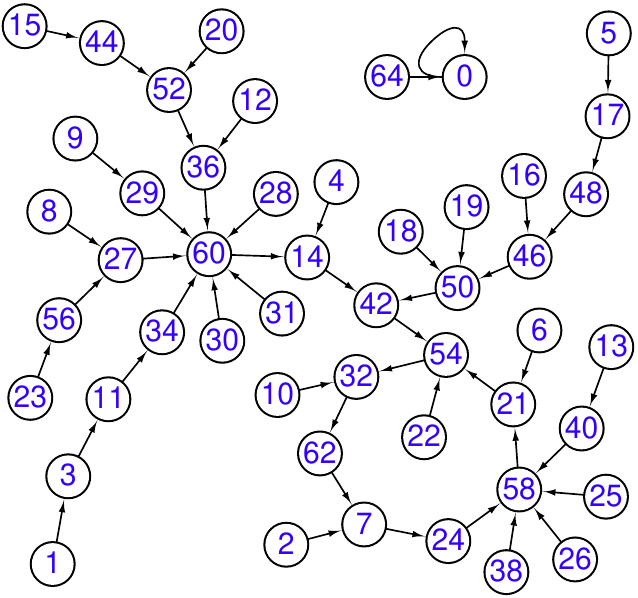}
			a)
		\end{minipage}  \hspace{\figsep}
		\begin{minipage}{\TwoImW}
			\centering
			\includegraphics[width=\TwoImW]{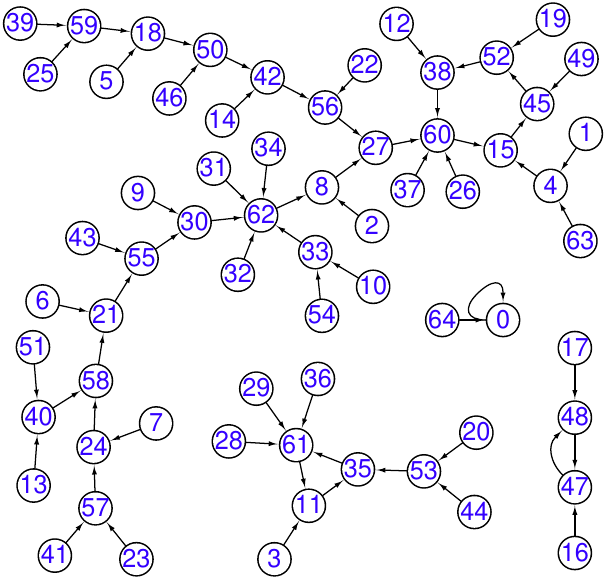}
			b)
		\end{minipage}
		\caption{SMN of the Logistic map with $\mu=62/2^4$:
			a) 8-bit floating-point domain with ($l=3$, $m=4$);
			b) 6-bit fixed-point precision and round quantization.}
		\label{fig:networkLogistic8and6bits}
	\end{figure}

\begin{figure}[!htb]
		\centering
		\begin{minipage}{0.9\TwoImW}
			\centering
			\includegraphics[width=0.9\TwoImW]{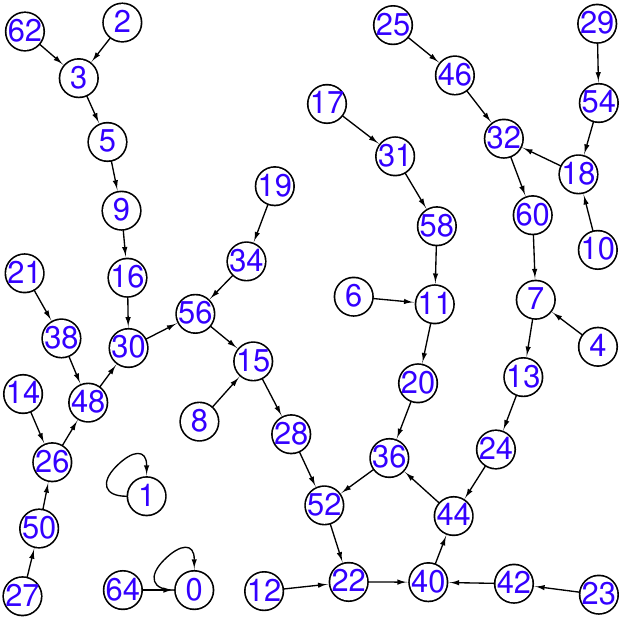}
			a)
		\end{minipage}\hspace{\figsep}
		\begin{minipage}{\TwoImW}
			\centering
			\includegraphics[width=\TwoImW]{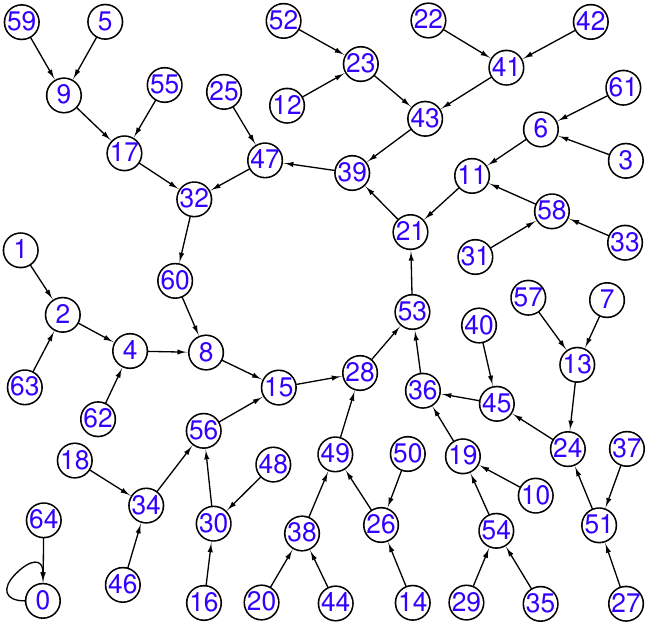}
			b)
		\end{minipage}
		\caption{SMN of the Tent map with $\mu=15/2^4$:
			a) 8-bit binary floating-point domain with ($l=3$, $m=4$);
			b) 6-bit fixed-point precision and round quantization.}
		\label{fig:networkTent8and6bits}
	\end{figure}

	The process can be verified by comparing Fig.~\ref{fig:networkLogistic8and6bits}a) and
	Fig.~\ref{fig:networkLogistic8and6bits}b) with Fig.~\ref{fig:networkTent8and6bits}a) and
	Fig.~\ref{fig:networkTent8and6bits}b), respectively. Table~\ref{tab:realdifftent} presents the
	differences between the two SMNs in the two arithmetic domains with $n=6$ and $(l, m)=(3, 4)$.
	
\begin{figure}[!htb]
		\centering
		\includegraphics[width=\Onefigwidth]{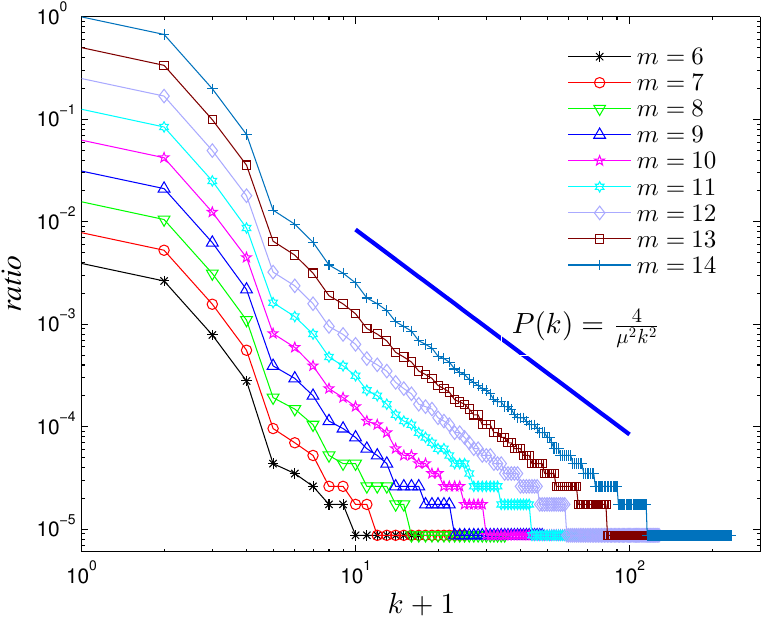}
		\caption{Cumulative in-degree distributions of the SMN of the Logistic map in various floating-point domains,
			where $l=4$, $m=6\sim 14$.}
		\label{fig:CumulativeInDegreeDistribution4}
	\end{figure}

\begin{figure}[!htb]
		\centering
		\includegraphics[width=\Onefigwidth]{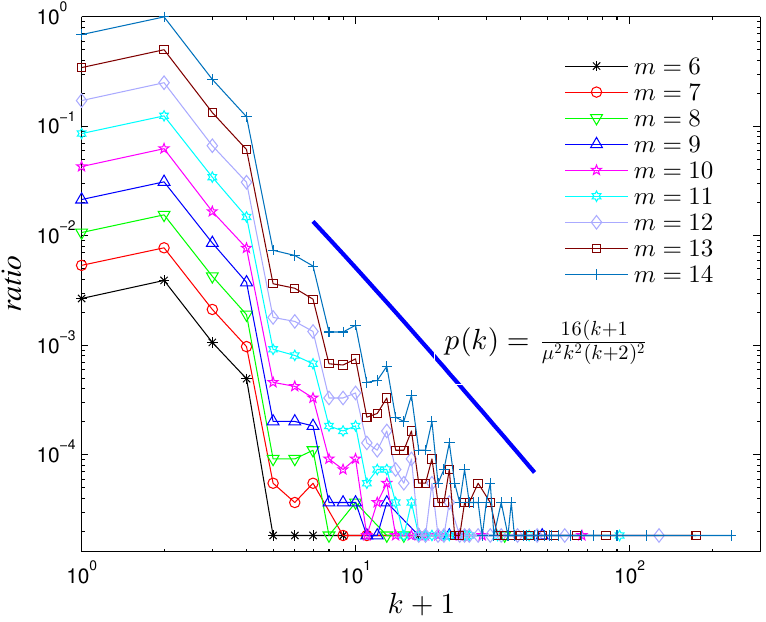}
		\caption{In-degree distributions of the SMN of the Logistic map in various floating-point domains,
			where $l=4$, $m=6\sim 14$.}
		\label{fig:InDegreeDistribution4}
	\end{figure}

	Referring to Theorems~\ref{theorem:logisticMap} and ~\ref{theorem:fixedFloat},
	one can conclude that the cumulative in-degree distribution and the in-degree distribution
	of the SMN of the Logistic map implemented in the floating-point arithmetic domain
	approximate that implemented in the corresponding fixed-point arithmetic domain.
	This is verified by comparing Fig.~\ref{fig:CumulativeInDegreeDistribution_u121_n_5_20}
	and Fig.~\ref{fig:CumulativeInDegreeDistribution4} with
	Fig.~\ref{fig:InDegreeDistribution_u121_n_5_20} and Fig.~\ref{fig:InDegreeDistribution4},
	respectively.
	
	\begin{table*}[!htb]
		\centering
		\caption{Differences between SMN implemented in the two arithmetic domains with $n=6$, $l=3$, $m=4$.}
		\begin{tabular}{c  *{11}{|c}}
			\hline
			$i$ & $F_{n}(i)$       & $f_{n}(i)-f_{l,m}(i)$ & $2^{-(m+1+j)}$ & $i$ & $F_{n}(i)$ &
			$f_{n}(i)-f_{l,m}(i)$  & $2^{-(m+1+j)}$ & $i$ & $F_{n}(i)$ & $f_{n}(i)-f_{l,m}(i)$  & $2^{-(m+1+j)}$\\ \hline
			0   & 0  & 0            & 0.015625 & 16 & 30 & 0            & 0.015625 & 32 & 60 & 0            & 0.03125  \\ \hline
			1   & 2  & 0.013671875  & 0.015625 & 17 & 32 & 0.013671875  & 0.03125  & 34 & 56 & 0.00390625   & 0.03125  \\ \hline
			2   & 4  & 0.01171875   & 0.015625 & 18 & 34 & 0.02734375   & 0.03125  & 36 & 53 & 0.0078125    & 0.03125  \\ \hline
			3   & 6  & 0.009765625  & 0.015625 & 19 & 36 & 0.025390625  & 0.03125  & 38 & 49 & 0.01171875   & 0.03125  \\ \hline
			4   & 8  & 0.0078125    & 0.015625 & 20 & 38 & 0.0234375    & 0.03125  & 40 & 45 & 0.015625     & 0.03125  \\ \hline
			5   & 9  & 0.005859375  & 0.015625 & 21 & 39 & 0.021484375  & 0.03125  & 42 & 41 & 0.01953125   & 0.03125  \\ \hline
			6   & 11 & 0.00390625   & 0.015625 & 22 & 41 & 0.01953125   & 0.03125  & 44 & 38 & 0.0234375    & 0.03125  \\ \hline
			7   & 13 & 0.001953125  & 0.015625 & 23 & 43 & 0.017578125  & 0.03125  & 46 & 34 & 0.02734375   & 0.03125  \\ \hline
			8   & 15 & 0            & 0.015625 & 24 & 45 & 0.015625     & 0.03125  & 48 & 30 & 0            & 0.015625 \\ \hline
			9   & 17 & 0.013671875  & 0.015625 & 25 & 47 & 0.013671875  & 0.03125  & 50 & 26 & 0.00390625   & 0.015625 \\ \hline
			10  & 19 & 0.01171875   & 0.015625 & 26 & 49 & 0.01171875   & 0.03125  & 52 & 23 & 0.0078125    & 0.015625 \\ \hline
			11  & 21 & 0.009765625  & 0.015625 & 27 & 51 & 0.009765625  & 0.03125  & 54 & 19 & 0.01171875   & 0.015625 \\ \hline
			12  & 23 & 0.0078125    & 0.015625 & 28 & 53 & 0.0078125    & 0.03125  & 56 & 15 & 0            & 0.015625 \\ \hline
			13  & 24 & 0.005859375  & 0.015625 & 29 & 54 & 0.005859375  & 0.03125  & 58 & 11 & 0.00390625   & 0.015625 \\ \hline
			14  & 26 & 0.00390625   & 0.015625 & 30 & 56 & 0.00390625   & 0.03125  & 60 & 8  & 0.0078125    & 0.015625 \\ \hline
			15  & 28 & 0.001953125  & 0.015625 & 31 & 58 & 0.001953125  & 0.03125  & 62 & 4  & 0.01171875   & 0.015625 \\ \hline
		\end{tabular}
		\label{tab:realdifftent}
	\end{table*}
	
	\begin{figure*}[!htb]
		\centering
		\begin{minipage}{\FiveImW}
			\centering
			\includegraphics[width=\FiveImW]{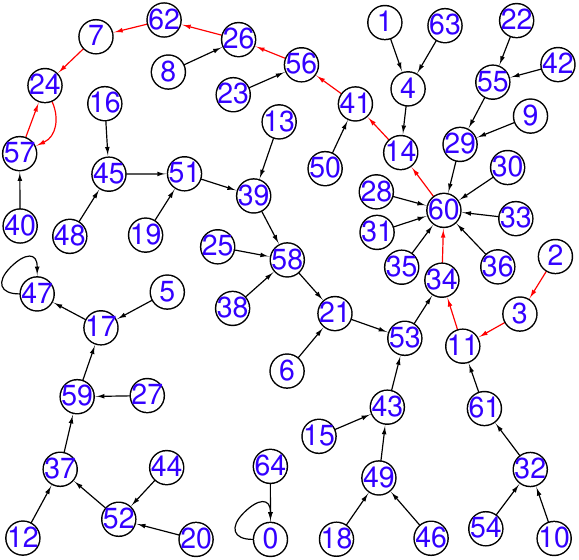}
			a)
		\end{minipage}
		\begin{minipage}{\FiveImW}
			\centering
			\includegraphics[width=\FiveImW]{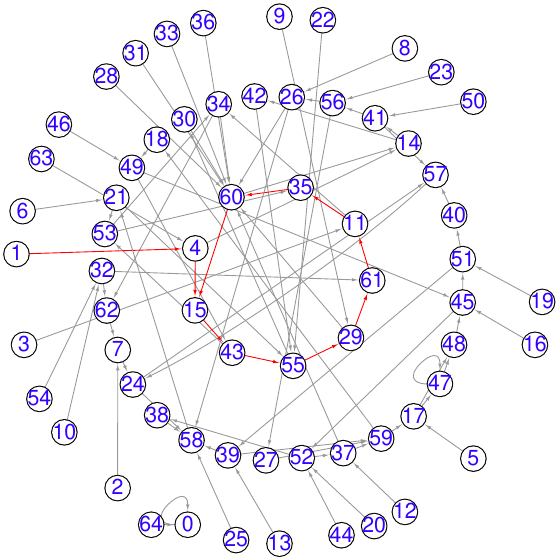}
			b)
		\end{minipage}
		\begin{minipage}{\FiveImW}
			\centering
			\includegraphics[width=\FiveImW]{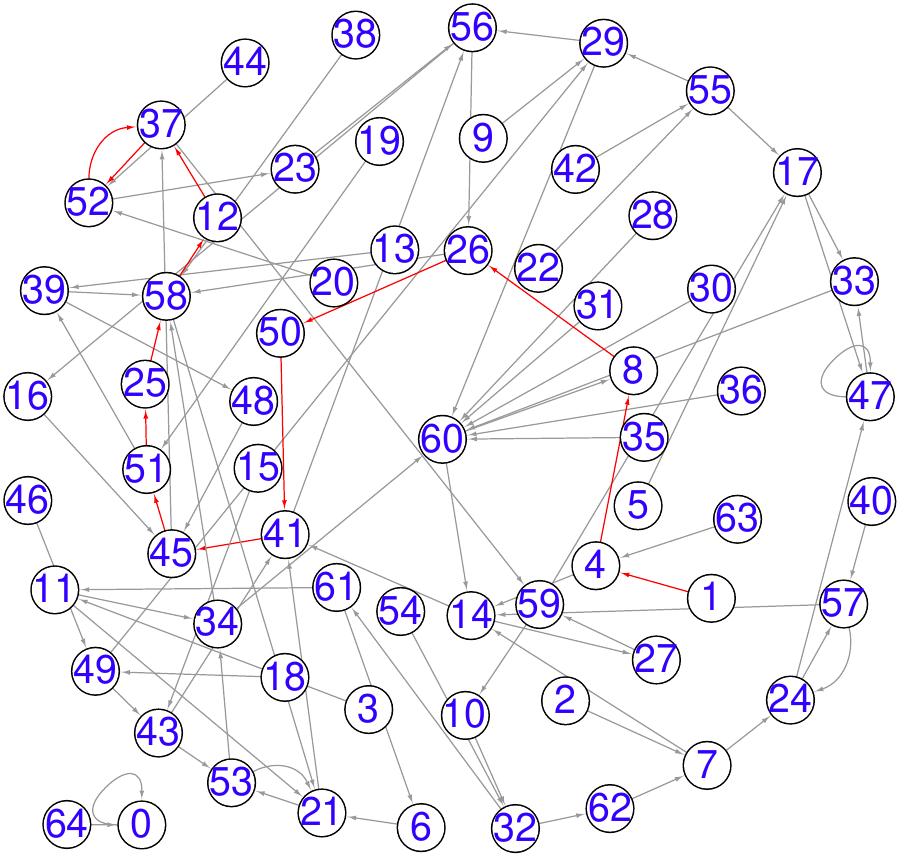}
			c)
		\end{minipage}
		\begin{minipage}{\FiveImW}
			\centering
			\includegraphics[width=\FiveImW]{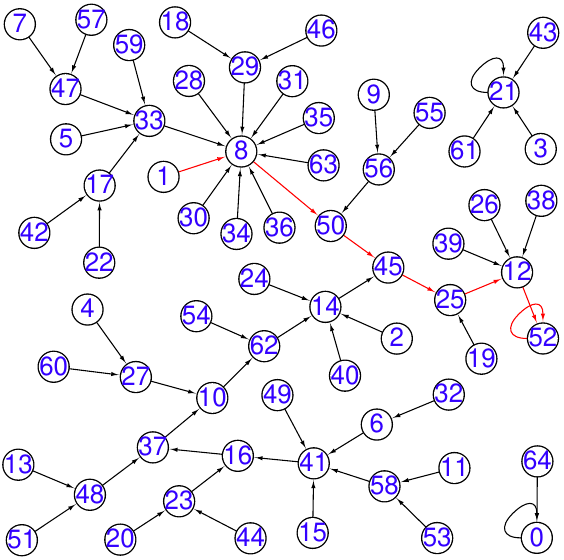}
			d)
		\end{minipage}
		\caption{Results of enhancing SMN shown in Fig.~\ref{fig:networkLogistic5and6bits}b)
			with various methods:
			a) perturbing states with the method in \cite{tao1998perturbance};
			b) perturbing control parameter $\mu$ from $121/2^5$ to $62/2^4$;
			c) switching with SMN of the Tent map with parameter $\mu=31/2^5$ alternately;
			d) cascading with SMN of the Tent map with parameter $\mu=31/2^5$.}
		\label{fig:networkMap6bits}
	\end{figure*}
	
	\section{Testing the randomness of various PRNGs based on iterating a chaotic map via SMN}
	\label{sec:applyNetwork}
	
	This section demonstrates that SMN can be used to classify the structures of various PRNGs based on iterating a chaotic map.
	Furthermore, it can work as a coarse visual tool for evaluating their randomness levels, as a complement to all kinds of test indexes enclosed
	in the standard test suites, e.g. NIST SP 800-22 \cite{Rukhin:TestPRNG:NIST} and
	TestU01 \cite{Pierre:TestU01:TMS07}. For this purpose, the various methods for chaos-based PRNG are classified into
	the following six categories by scrutinizing the SMNs of enhanced Logistic maps:
	\begin{itemize}[%
		\setlength{\labelwidth}{\widthof{\textbullet}}%
		\setlength{\labelsep}{3pt}%
		\setlength{\IEEElabelindent}{0pt}%
		\IEEEiedlabeljustifyl
		]
		\item \textit{Selecting state and control parameters}
		
		By observing Fig.~\ref{fig:networkLogistic5and6bits}, one can see that the nodes labeled ``0`` and
		``$``2^n$'' are pathological seeds, therefore they should be excluded from any PRNG. More similar nodes
		can be found from Fig.~\ref{fig:networkLogistic5and6bits}c), which are very difficult to be identified by a
		randomness test suite.  In \cite{Addabbo:Tent:ITIM2006}, it was claimed that the discrete Tent
		map can achieve a satisfactory trade-off among the period lengths, the statistical characteristics of
		the generated bit sequences, and the complexity of hardware implementations in the fixed-point
		domain. As the control parameter is the sole factor in the SMN of the Tent map under the given
		implementation environment, now one can see that in \cite{heidari1994chaotic} it only selects the desired SMN by
		choosing some control parameters.
		
		\item \textit{Increasing the arithmetic precision}
		
		As discussed in \cite{lin1991chaos}, increasing the arithmetic precision can substantially enlarge
		the average length of the orbit of an SMN and enhance its complexity. However, such enhancing method cannot change its overall structure, as can be observed from Fig.~\ref{fig:networkLogistic5and6bits}. Figure~2 in \cite{Persohn:AnalyzeLogistic:CSF12}
		also confirms that increasing the precision does not always enlarge the average path period of an SMN.
		In addition, exhaustively searching smaller-scale data incurred by a lower arithmetic precision,
		e.g. binary16, may also discover the rules found from the big data generated by a higher-precision
		as in \cite{Persohn:AnalyzeLogistic:CSF12}.
		
		\item \textit{Perturbing states}
		
		Essentially, perturbing the state is to jump from a walk path in an SMN to another (namely, to rewire
		the linking edges of an SMN)
		\cite{tao1998perturbance,heidari1994chaotic,Chen:random:CASII2010,LiCY:PRNS:VLSI2012}.
		To show the effect of this kind of methods, the SMN is perturbed as shown in Fig.~\ref{fig:networkLogistic5and6bits}b),
		by the method given in \cite{tao1998perturbance}. The result is presented in Fig.~\ref{fig:networkMap6bits}a),
		where the perturbation is performed by bit-wise XOR of the three least significant bits of the
		mapping value and the perturbing bit sequence $(100)_{2}$. From Fig.~\ref{fig:networkMap6bits}a),
		one can see that the orbit starting from some states, especially that in connected components of
		small sizes, remains unchanged. A cycle may become even shorter after the states are perturbed.
		Generally, the enhancing methods based on feed-back control proposed in
		\cite{addabbo2006feedback,HPhu:DCS:SMCS2015} can be considered as perturbing nodes of the corresponding SMN
		adaptively.
		
		\item \textit{Perturbing the control parameters}
		
		Perturbing the control parameters is to walk from a path of an SMN corresponding to one control parameter
		to that corresponding to another one with a timely jump. So, this kind of methods is actually to cascade
		multiple SMNs generated by the same chaotic map \cite[Sec. 4]{vcernak1996digital}. To visualize this
		strategy, SMN cascades are shown in Fig.~\ref{fig:networkLogistic5and6bits}b), along with that in
		Fig.~\ref{fig:networkLogistic8and6bits}b), and the results are presented in Fig.~\ref{fig:networkMap6bits}b).
		
		% As discussed in \cite{Uehara:logistic:ITA}, the effect on SMN by a tiny change of the control parameter may be canceled by quantization operations.
		
		\item \textit{Switching among multiple chaotic maps}
		
		In each iteration, the chaotic map is switched from one candidate to another, so as to generate
		the next state \cite{nagaraj2008increasing,YCZhou:Switching:TCASI2014}.
		From the viewpoint of SMN, the obtained orbit is to walk on every SMN for one step and then jump
		to another SMN, depending on the switching operation. As one state may be operated by different
		chaotic maps, the out-degrees of some nodes of the final SMN may be larger than 1 but bounded
		by the number of available chaotic maps. A demo on switching between the Logistic map and the
		Tent map is shown in Fig.~\ref{fig:networkMap6bits}c).
		
		\item \textit{Cascading among multiple chaotic maps}
		
		This kind of methods is to cascade some walks on multiple SMNs into one walk \cite{Hua:model:CAS17}.
		The method used in \cite{heidari1994chaotic} is an extreme case, where the outputs of one chaotic
		map are used to select the start of the path in the SMN of another chaotic map. Although two chaotic
		maps were used, the finally obtained SMN is not updated. A demo of the SMN, on cascading two
		chaotic maps shown in Fig.~\ref{fig:networkLogistic5and6bits}b) and Fig.~\ref{fig:networkTent5and6bits}
		respectively, is depicted in Fig.~\ref{fig:networkMap6bits}d). The isolated but connected components
		in the original SMN can be connected together into the final cascaded SMN once they own one pair of nodes
		connected in any SMN.
	\end{itemize}
	
	As above enhancing methods can be considered to make the dynamical properties of an existing chaotic map become more complex,
	SMN can also be used to evaluate its dynamical complexity as did in \cite{WangQX:HDDCS:TCAS2016}.
	
	% From the above discussion, one can easily compare the randomness performance of various enhancement methods for PRNG from the viewpoint of SMN.
	
	\section{Conclusions}
	
	This paper has studied the dynamical properties of digital chaotic maps with the methodology
	of complex networks. Some subtle properties of the state-mapping networks of the Logistic
	map and the Tent map have been revealed, offering a panorama with both microscopic and
	macroscopic structures of the network. It has been demonstrated that the state-mapping
	network of a digital map in a small-precision digital domain can work as an efficient tool
	for classifying its structure and coarsely verify its randomness. This analysis can be further extended to
	higher-dimensional chaotic systems. Achievements notwithstanding, more properties and
	applications of the state-mapping network framework of various chaotic maps call for further
	exploration in the near future.
	
	\bibliographystyle{IEEEtran_doi}
	\bibliography{PRNS_Chaos}
	
	%\iffalse
	%%%%%%%%%%%%%%%%%%%%%%%%%%%%%%%%%%%%%%%%%%%%%%%%%%%%%%
	\graphicspath{{author-figures-pdf/}}
	
	%\vspace{-30mm}
	
	% \newpage
	
	\begin{IEEEbiography}[{\includegraphics[width=1.1in, height=1.25in,clip,keepaspectratio]{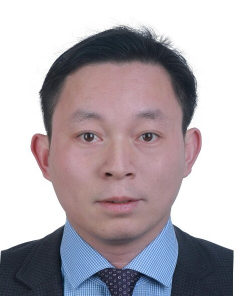}}]{Chengqing Li}(M'07--SM'13) received his M.Sc.\ degree in applied mathematics from Zhejiang University, China in 2005 and his Ph.D.\ degree in electronic engineering from City University of Hong Kong in 2008. Thereafter, he worked as a Postdoctoral Fellow at The Hong Kong Polytechnic University till September 2010. Then, he worked at the College of Information Engineering, Xiangtan University, China. From April 2013 to July 2014, he worked at the University of Konstanz, Germany, under the support of the Alexander von Humboldt Foundation.  Since April 2018, he has been working with the College of Computer Science and Electronic Engineering, Hunan University, China as a full professor.
		% He is serving as an associate editor for the International Journal of Bifurcation and Chaos.
		
		Prof.\ Li focuses on dynamics analysis of digital chaotic systems and their applications in multimedia security. He has published about fifty papers on the focal subject in the past 13 years, receiving more than 2500 citations with h-index 28.
	\end{IEEEbiography}
	
	\vspace{-18mm}
	
	\begin{IEEEbiography}[{\includegraphics[width=1.1in, height=1.25in,clip,keepaspectratio]{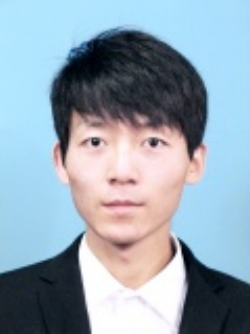}}]{Bingbing Feng} obtained his B.Sc.\ degree and M.Sc.\ degree both in computer science from the College of Information Engineering, Xiangtan University, China in 2015 and 2018, respectively.
		
		His research interests include complex networks and nonlinear dynamics.
	\end{IEEEbiography}
	
	\vspace{-14mm}
	
	\begin{IEEEbiography}[{\includegraphics[width=1.1in, height=1.25in,clip,keepaspectratio]{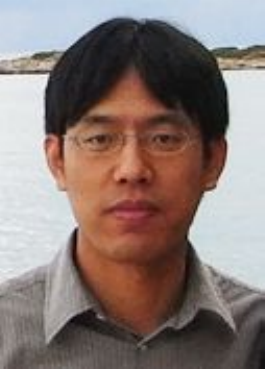}}]{Shujun Li} (M'08--SM'12) received the B.E.\ degree in information science and engineering, and the Ph.D.\ degree in information and communication engineering from Xi'an Jiaotong University, China, in 1997 and 2003, respectively. From 2003 to 2007, he was a Post-Doctoral Research Assistant with the City University of Hong Kong, and then a Post-Doctoral Fellow with the Hong Kong Polytechnic University. From 2007 to 2008, he was conducting visiting research at the FernUniversit\``{a}t, Hagen, Germany, as a Humboldt Research Fellow. From 2008 to 2011, he was a Zukunftskolleg Fellow with the Universit\``{a}t Konstanz, Germany. In 2011, he joined University of Surrey, UK, initially as a Senior Lecturer and then was promoted to Reader in 2017. Since November 2017, he has been a Professor of Cyber Security at the University of Kent, UK, and directing the Kent Interdisciplinary Research Centre in Cyber Security (KirCCS), a UK government recognized Academic Centre of Excellence in Cyber Security Research (ACE-CSR).
		
		Prof.\ Li's current research interests mainly focus on interplays between several interdisciplinary research areas including cyber security and privacy, cybercrime, human factors, multimedia computing, and digital chaos. He is a Fellow of the BCS -- The Chartered Institute for IT and the Vice President for Internal Communications \& Public Relations of the ABCP (Association of British Chinese Professors).
	\end{IEEEbiography}
	%\enlargethispage{-9.5cm}
	\vspace{-10mm}
	
	\begin{IEEEbiography}[{\includegraphics[width=1.1in, height=1.25in,clip,keepaspectratio]{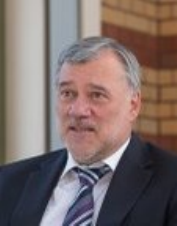}}]{J\"urgen Kurths} received the Ph.D.\ degree from the GDR Academy of Sciences, Berlin, Germany, in 1983. He was a Full Professor at the University of Potsdam from 1994 to 2008. He has been a Professor of nonlinear dynamics at Humboldt University of Berlin, Berlin, and the Chair of the research domain Transdisciplinary Concepts of the Potsdam Institute for Climate Impact Research, Potsdam, Germany, since 2008. He has authored or coauthored more than 500 papers that are cited more than 18,000 times (h-index: 57). He became a member of the Academy of Europe in 2010 and of the Macedonian Academy of Sciences and Arts in 2012. He is the Editor-in-chief of CHAOS.
		
		Prof.\ Kurths' primary research interests include synchronization, complex networks, and time-series analysis and their applications.
	\end{IEEEbiography}
	
	\vspace{-11mm}
	
	\begin{IEEEbiography}[{\includegraphics[width=1.1in, height=1.25in, clip, keepaspectratio]{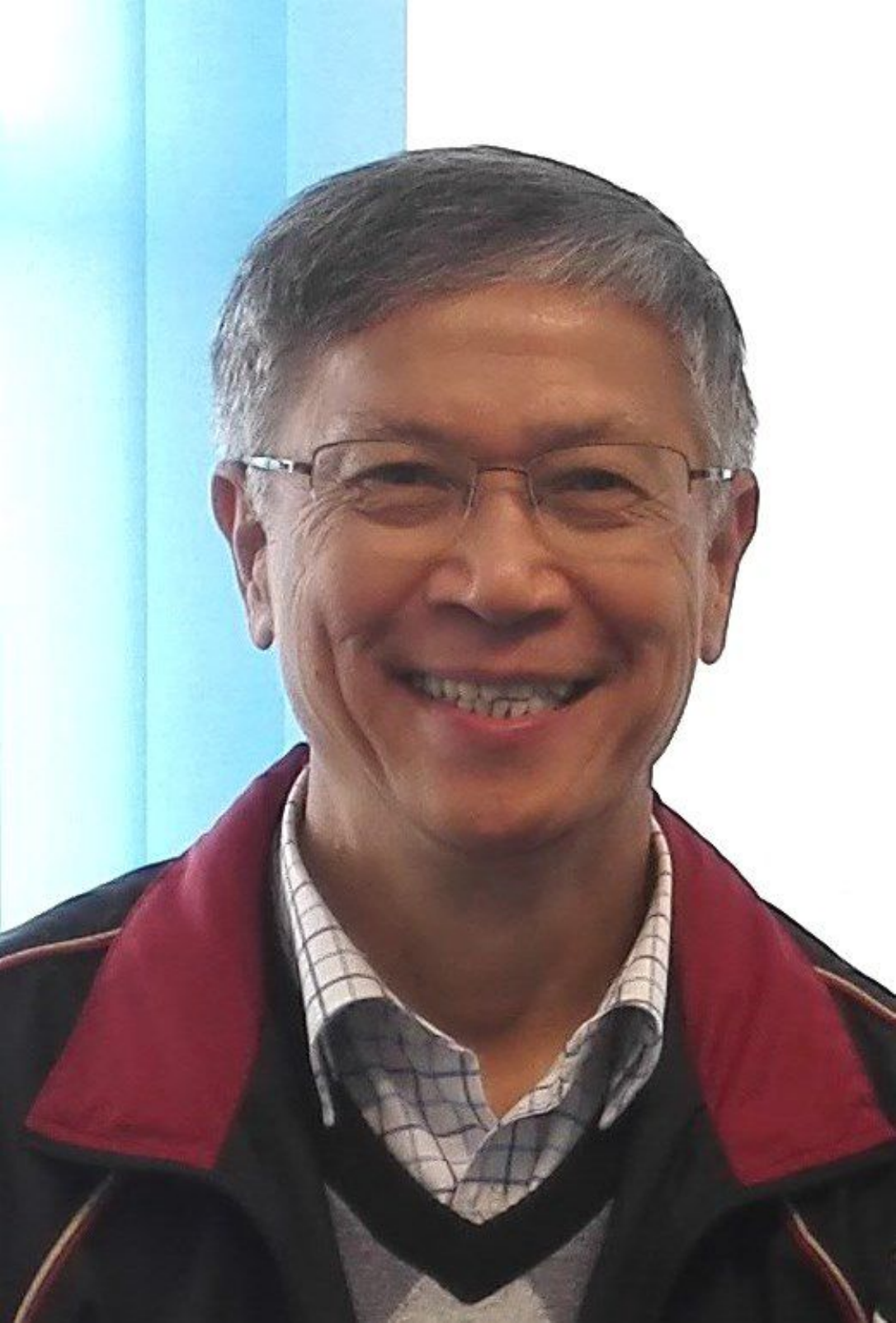}}]{Guanrong Chen} (M'89--SM'92--F'97--LF'19) received the MSc.\ degree in computer science from Sun Yat-sen University, Guangzhou, China in 1981 and the Ph.D.\ degree in applied mathematics from Texas A\&M University, TX, USA in 1987. He has been a Chair Professor and the Director of the Centre for Chaos and Complex Networks, City University of Hong Kong, Hong Kong since 2000, prior to that he was a tenured Full Professor with the University of Houston, Houston, TX, USA.
		
		Prof.\ Chen was a recipient of the 2011 Euler Gold Medal, Russia, and Highly Cited Researcher in Engineering named by Thomson Reuters, and conferred Honorary Doctorate by the Saint Petersburg State University, Russia in 2011 and by the University of Le Havre, Normandie, France in 2014. He is a member of the Academy of Europe and a fellow of The World Academy of Sciences.
	\end{IEEEbiography}
	
	%\enlargethispage{-9.5cm}
	%\enlargethispage{\baselineskip}
	%\fi
	%\vfill
\end{document}